\documentclass[reprint,aps,prd,superscriptaddress,nofootinbib,longbibliography]{revtex4-2}

\usepackage{amsmath,amssymb}
\usepackage{graphicx}
\usepackage{bm}
\usepackage{mathrsfs}
\usepackage[colorlinks=true,allcolors=blue]{hyperref}
\usepackage{orcidlink}

\begin{document}

\title{Charged particles and QPO constraints in tidal charge black hole magnetosphere sourced by a current loop}

\author{Ozodbek~Abdurakhmonov\orcidlink{0009-0001-0439-6132}} \email{ozodbek992606@gmail.com} 
\affiliation{Institute of Fundamental and Applied Research, National Research University TIIAME, Kori Niyoziy 39, Tashkent 100000, Uzbekistan}

\author{Javlon~Rayimbaev}
\email{javlonrayimbaev6@gmail.com}
\affiliation{Institute of Theoretical Physics, National University of Uzbekistan, Tashkent 100174, Uzbekistan}
\affiliation{University of Tashkent for Applied Sciences, Str. Gavhar 1, Tashkent 100149, Uzbekistan}

\author{Inomjon~Ibragimov \orcidlink{0009-0009-6979-6354}
}
\email{i.ibragimov@kiut.uz}
\affiliation{Kimyo International University in Tashkent, Shota Rustaveli Street 156, Tashkent 100121, Uzbekistan}

\author{Sherzod Djumanov}
\email{djumanov@inp.uz}
\affiliation{Tashkent State Technical University, Tashkent 100095, Uzbekistan}

\author{Nuraliev~Farhod\orcidlink{0009-0009-1711-2088}}
\email{nuraliyevf@mail.ru}
\affiliation{Tashkent International University, Little Ring Road 7, Tashkent 100115, Uzbekistan}

\author{Ilkhomjon Makhmudov\orcidlink{0000-0002-6026-5292}}
\email{ilhom_makhmudov@mail.ru}
\affiliation{Research Institute of Irrigation and Water Problems, Tashkent, 100187, Uzbekistan}

\date{\today}

\begin{abstract}
We investigate the dynamics of charged particles in the dipolar electromagnetic field generated by a stationary current loop around a static black hole in the Randall--Sundrum braneworld model. The current loop is located at the innermost stable circular orbit (ISCO) of the tidal charged black hole, and the azimuthal electromagnetic four-potential is obtained as an exact analytical solution of the Maxwell equations in the curved background. Starting from the Hamilton--Jacobi equation, we derive the effective potential for charged particle motion in both the interior and exterior regions of the current loop, and analyze the dependence of the ISCO radius, angular momentum, energy, and orbital frequency on the tidal charge parameter $b$ and the magnetic coupling parameter $\omega$. We further compute three fundamental frequencies and show that the dipolar loop field revives a nonzero nodal precession frequency even though the background is static and non-rotating. Using these frequencies we construct five twin-peak QPO models (RP, WD, and ER2--ER4) and perform a Markov chain Monte Carlo (MCMC) analysis to constrain the braneworld parameters $(b,\omega)$ from the observed twin-peak QPO frequencies of six black hole sources spanning three mass regimes: GRO~J1655--40, XTE~J1550--564, GRS~1915$+$105, H~1743$+$322, M82~X-1, and Sgr~A$^{*}$. Adopting the epicyclic-resonance ER4 model as the representative framework, we find moderate, non-negligible tidal charges $b\simeq0.3$--$1.0$, while the magnetic coupling remains consistent with zero, $|\omega|\lesssim0.05$. Our results therefore place upper bounds on the tidal charge and on the magnetic coupling strength from astrophysical QPO data.
\end{abstract}

\maketitle

\section{Introduction}
\label{sec:intro}

The electromagnetic field in the vicinity of black holes plays a central role in understanding the observational signatures of compact objects such as quasars, blazars, and accreting X-ray sources~\cite{Petterson1974,Wald1974,Piotrovich2010}. In realistic astrophysical environments, black holes are surrounded by magnetized plasma, and the resulting field configuration governs the dynamics of charged particles, the structure of accretion flows, and the emission of high-energy radiation~\cite{Wald1974,AlZahrani2013,Tursunov2018}.
Because the no-hair theorem forbids an isolated black hole from carrying an intrinsic electromagnetic field, any such field must be sourced externally---most naturally by the current-carrying plasma of the surrounding disk.

The dipolar electromagnetic field of a stationary current loop around a Schwarzschild black hole was first obtained by Petterson~\cite{Petterson1974} and subsequently generalized to the Kerr spacetime~\cite{Petterson1975}. Black holes immersed in asymptotically uniform magnetic fields have been studied extensively since Wald~\cite{Wald1974,Piotrovich2010}, and researchers have analyzed the dipolar field configuration of magnetized compact stars in both Newtonian and general relativistic frameworks~\cite{Deutsch1955,Ginzburg1964,Rezzolla2001a,Rezzolla2001b}.
A broad body of work has shown that even weak external fields strongly modify charged-particle motion, shifting characteristic radii such as the ISCO and reshaping the phase-space structure of the orbits~\cite{AlZahrani2013,Igata2012,Kopacek2014,Panis2019,Stuchlik2020,KhanChen2023}.

The braneworld scenario proposed by Randall and Sundrum~\cite{Randall1999} provides a compelling higher-dimensional framework in which matter is confined to a four-dimensional brane while gravity propagates in the bulk. The static, spherically symmetric vacuum solution on the brane was obtained by Dadhich \emph{et al.}~\cite{Dadhich2000}; it has the same functional form as the Reissner--Nordstr\"om metric, with a tidal charge $Q_*$ replacing the square of the electric charge. Crucially, the tidal charge may be negative, $Q_*<0$, which \emph{strengthens} the gravitational field and has no analogue in general relativity. The electromagnetic field of a current loop around a tidally charged black hole was constructed by Turimov~\cite{Turimov2018}, who obtained exact analytical expressions for the dipolar vector potential in both the interior and exterior regions of the loop. Braneworld effects on particle motion and electromagnetic fields around tidal charged black holes have been investigated in Refs.~\cite{Aliev2005,Abdujabbarov2010,Stuchlik2017,Schee2009a,Schee2009b}, while the magnetized relativistic star in the braneworld has been treated in Refs.~\cite{Turimov2017,Ahmedov2008,Morozova2011}.

Quasi-periodic oscillations (QPOs) observed in the X-ray power spectra of accreting compact objects provide one of the most direct probes of the strong-field regime of gravity~\cite{Stella1998,Stella1999,Belloni2012,Remillard2002}. The relativistic precession model (RPM)~\cite{Stella1999,Motta2014} relates the observed twin-peak frequencies to the fundamental orbital and epicyclic frequencies of test-particle motion, while epicyclic-resonance and warped-disc models attribute the twin peaks to nonlinear resonances between the radial and vertical oscillation modes of the disk~\cite{AbramowiczKluzniak2001,Torok2005,IngramMotta2019}. QPO data have been widely used to constrain modified-gravity parameters and to test the no-hair theorem~\cite{Bambi2012,Johannsen2011,Bambi2014}, and recent analyses employ Bayesian inference and MCMC sampling to extract black-hole parameters directly from the twin-peak frequencies~\cite{Bambi2014,ForemanMackey2013}.

In the present work, we extend the analysis of Ref.~\cite{Turimov2018} to \emph{charged} particles moving in the dipolar field of a current loop around a tidal charged black hole. Working from the Hamilton--Jacobi equation, we derive the charged-particle effective potential, analyze the circular orbits and the ISCO as functions of both the tidal charge $b$ and the magnetic coupling $\omega$, compute the three fundamental frequencies of the configuration, and finally confront five twin-peak QPO models with the observed frequencies of six sources through an MCMC analysis~\cite{Babar2016,Turimov2020,Kurbonov2025}.

The paper is organized as follows. In Sec.~\ref{sec:spacetime} we review the tidal charged black hole spacetime and neutral particle motion. In Sec.~\ref{sec:emfield} we present the electromagnetic four-potential of the current loop and derive the charged-particle effective potential. In Sec.~\ref{sec:circular} we analyze the circular orbits and ISCO properties. In Sec.~\ref{sec:frequencies} we compute the fundamental (Keplerian and epicyclic) frequencies. In Sec.~\ref{sec:qpo} we build the twin-peak QPO models, and in Sec.~\ref{sec:mcmc} we perform MCMC parameter estimation. We summarize our results in Sec.~\ref{sec:conclusion}. Throughout we use geometrized units $G=c=1$ and the signature $(-,+,+,+)$.

\section{Tidal charged black hole spacetime}
\label{sec:spacetime}

In the Randall--Sundrum braneworld model~\cite{Randall1999}, the static, spherically symmetric vacuum solution on the brane is described by the line element~\cite{Dadhich2000}
\begin{equation}
ds^2 = -N_*^2\,dt^2 + \frac{dr^2}{N_*^2} + r^2(d\theta^2 + \sin^2\theta\,d\varphi^2)\,,
\label{eq:metric}
\end{equation}
where the lapse function is
\begin{equation}
N_*^2(r) = 1 - \frac{2M}{r} + \frac{Q_*}{r^2}\,,
\label{eq:lapse}
\end{equation}
with $M$ the total mass and $Q_*$ the tidal charge arising from the projection of the five-dimensional Weyl tensor onto the brane~\cite{Dadhich2000}. The tidal charge is negatively defined, $Q_*<0$, and we introduce the dimensionless positive parameter
\begin{equation}
b = \frac{|Q_*|}{M^2} > 0\,,
\label{eq:bdef}
\end{equation}
so that
\begin{equation}
N_*^2(r) = 1 - \frac{2M}{r} - \frac{M^2 b}{r^2}\,.
\label{eq:lapseb}
\end{equation}
The event horizon is located at
\begin{equation}
r_+ = M\left(1 + \sqrt{1+b}\right)\,.
\label{eq:horizon}
\end{equation}

\begin{figure}[t]
\centering
\includegraphics[width=0.9\linewidth]{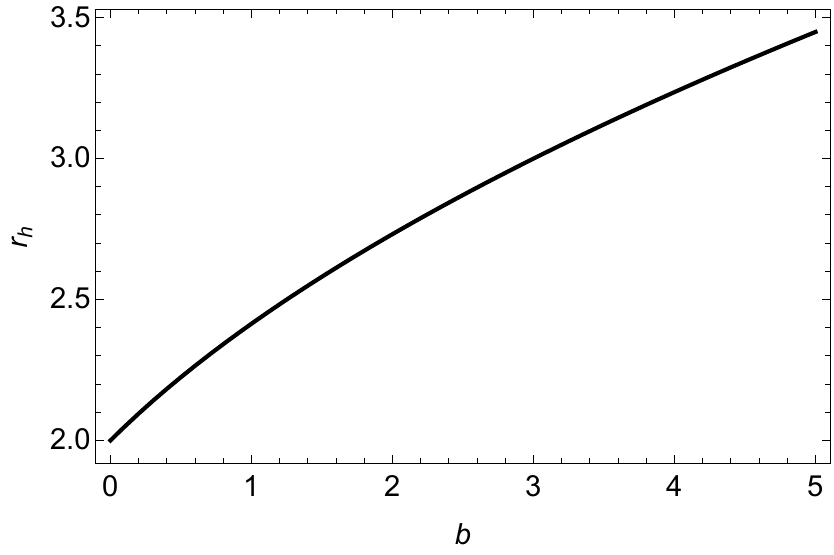}
\caption{The event horizon radius $r_h$ as a function of the tidal charge parameter $b$. At $b=0$ the Schwarzschild limit $r_h = 2M$ is recovered.}
\label{fig:horizon_b}
\end{figure}

The horizon radius $r_h = M(1+\sqrt{1+b})$ grows monotonically with $b$ (Fig.~\ref{fig:horizon_b}), recovering the Schwarzschild value $r_h=2M$ at $b=0$; the negative tidal charge thus enlarges the horizon relative to Schwarzschild, in line with its gravitationally attractive character~\cite{Dadhich2000,Stuchlik2017}.

\subsection{Particle motion and ISCO}

The motion of a test particle of mass $m$ is governed by the Hamilton--Jacobi equation. For equatorial motion ($\theta=\pi/2$) the radial equation reads~\cite{Turimov2018,Schee2009a}
\begin{equation}
\left(\frac{dr}{d\tau}\right)^2 = \mathcal{E}^2 - V_{\rm eff}^{(0)}(r)\,,
\label{eq:radial_neutral}
\end{equation}
where $\mathcal{E}=E/m$ and $\mathcal{L}=L/m$ are the specific energy and angular momentum, and the effective potential for a neutral particle is
\begin{equation}
V_{\rm eff}^{(0)}(r) = N_*^2(r)\left(1 + \frac{\mathcal{L}^2}{r^2}\right)\,.
\label{eq:veff_neutral}
\end{equation}
The circular orbit conditions $V_{\rm eff}^{(0)}=\mathcal{E}^2$ and
$dV_{\rm eff}^{(0)}/dr=0$ yield~\cite{Turimov2018}
\begin{equation}
\mathcal{E}^2 = \frac{(r^2 - 2Mr - M^2 b)^2}{r^2(r^2 - 3Mr - 2M^2 b)}\,,
\label{eq:E_neutral}
\end{equation}
\begin{equation}
\mathcal{L}^2 = \frac{r^2 M(r + Mb)}{r^2 - 3Mr - 2M^2 b}\,.
\label{eq:L_neutral}
\end{equation}
The ISCO radius follows from the additional condition
$d^2V_{\rm eff}^{(0)}/dr^2=0$~\cite{Turimov2018,Stuchlik2017},
\begin{equation}
r_{\rm ISCO} = M\left(2 + \eta(b) + \frac{4 + 3b}{\eta(b)}\right)\,,
\label{eq:risco}
\end{equation}
with
\begin{equation}
\eta(b) = \left(8 + 9b + 2b^2 - b\sqrt{(1+b)(5+4b)}\right)^{1/3}\,.
\label{eq:eta}
\end{equation}
In the Schwarzschild limit $b=0$ one recovers $r_{\rm ISCO}=6M$.

\begin{figure}[t]
\centering
\includegraphics[width=0.9\linewidth]{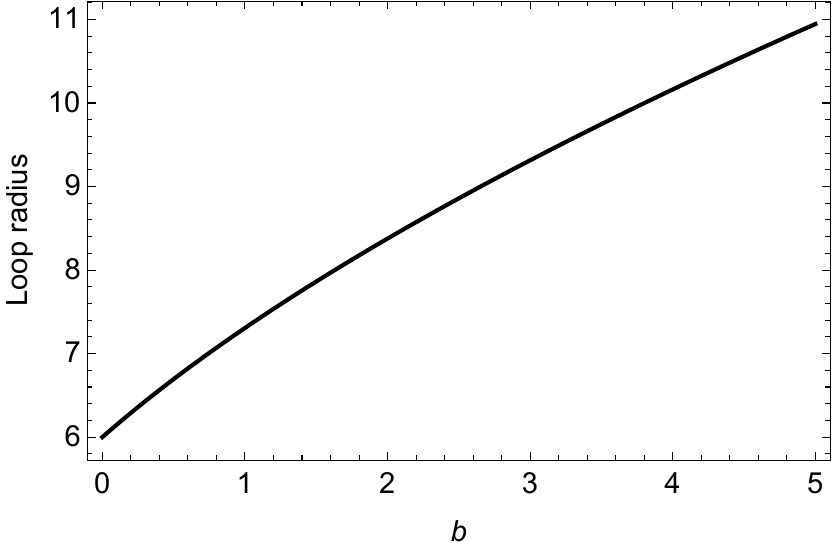}
\caption{The ISCO radius $r_{\rm ISCO}$ (in units of $M$) as a function of the tidal charge parameter $b$. At $b=0$ the Schwarzschild value $r_{\rm ISCO}=6M$ is recovered.}
\label{fig:isco_b}
\end{figure}

The current loop is placed at the neutral ISCO, $r_0=r_{\rm ISCO}(b)$, which increases monotonically with $b$ (Fig.~\ref{fig:isco_b}); for example $r_{\rm ISCO}\simeq7.31M$ at $b=1$ and $\simeq8.38M$ at $b=2$, recovering $6M$ at $b=0$. The outward migration of the ISCO is the geometric signature of the enhanced gravitational attraction produced by the negative tidal charge~\cite{Stuchlik2017,Schee2009a}.

\section{Electromagnetic field and charged particle effective potential}
\label{sec:emfield}

\subsection{Dipolar electromagnetic field of the current loop}

We consider a stationary current loop generated by charged matter orbiting the black hole at the ISCO radius $r_0=r_{\rm ISCO}(b)$~\cite{Petterson1974,Turimov2018}. For the axially symmetric, stationary field, the Maxwell equations in the spacetime~\eqref{eq:metric} reduce to~\cite{Turimov2018}
\begin{align}
\frac{1}{\sin^2\theta}\frac{\partial}{\partial r}
\left(N_*^2\frac{\partial A_\varphi}{\partial r}\right)
&+ \frac{1}{r^2\sin\theta}\frac{\partial}{\partial\theta}
\left(\frac{1}{\sin\theta}\frac{\partial A_\varphi}{\partial\theta}\right) \nonumber\\
&= -4\pi I N_*\,\delta(r - r_0)\,\delta(\cos\theta)\,,
\label{eq:maxwell}
\end{align}
where $I$ is the total electric current in the loop and $A_\varphi$ is the azimuthal component of the four-potential. Seeking a separable solution $A_\varphi(r,\theta)=R(r)\,T(\theta)$ in the source-free region, the angular and radial parts decouple into
\begin{align}
\sin\theta\,\frac{d}{d\theta}\!\left(\frac{1}{\sin\theta}\frac{dT_l}{d\theta}\right)
+(l+1)(l+2)\,T_l(\theta) &= 0\,, \label{eq:angular_eq}\\
\left[(r-M)^2 - M^2\varrho^2\right]R_l'' + 2M\!\left(1+\frac{Mb}{r}\right)R_l'
-(l+1)(l+2)R_l &= 0\,, \label{eq:radial_gen}
\end{align}
where $l$ is an integer multipole index. Restricting to the lowest (dipole) mode $l=0$, the
regular angular solution of Eq.~\eqref{eq:angular_eq} is
\begin{equation}
T(\theta) = \sin^2\theta\,,
\label{eq:angular}
\end{equation}
and the radial equation~\eqref{eq:radial_gen} becomes
\begin{equation}
\left[(r-M)^2 - M^2\varrho^2\right]R'' + 2M\left(1 + \frac{Mb}{r}\right)R' - 2R = 0\,,
\label{eq:radial_eq}
\end{equation}
with $\varrho=\sqrt{1+b}$. This equation admits two linearly independent solutions
\begin{equation}
U_0(r) = r^2 + M^2 b\,,
\label{eq:U0}
\end{equation}
\begin{align}
V_0(r) = \frac{1}{8\varrho^3 M^3}\bigg[&2M(r+M) \nonumber\\
&+ (r^2 + M^2 b)\ln\frac{r - M - M\varrho}{r - M + M\varrho}\bigg]\,,
\label{eq:V0}
\end{align}
of which $U_0$ is regular at the horizon but diverges at infinity, and $V_0$ is regular at infinity but diverges at the horizon. Imposing regularity in each region and continuity across the loop at $r=r_0$, the exact analytical potential is~\cite{Turimov2018}
\begin{widetext}
\begin{equation}
A_\varphi = -\frac{3\mu_*\sin^2\theta}{8\varrho^3 M^3}(r^2 + M^2 b) \times
\begin{cases}
\displaystyle\frac{2\varrho M(r_0 + 2M)}{r_0^2 + M^2 b}
+ \ln\dfrac{r_0 - M - M\varrho}{r_0 - M + M\varrho}\,, & r_+ \leq r \leq r_0\,,\\[12pt]
\displaystyle\frac{2\varrho M(r + 2M)}{r^2 + M^2 b}
+ \ln\dfrac{r - M - M\varrho}{r - M + M\varrho}\,, & r \geq r_0\,,
\end{cases}
\label{eq:Aphi_full}
\end{equation}
\end{widetext}
where the magnetic dipole moment is
\begin{equation}
\mu_* = \pi r_0^2\,I\,N_*(r_0)\left(1 + \frac{M^2 b}{r_0^2}\right)\,.
\label{eq:mu}
\end{equation}
For convenience, we define the bracket functions
\begin{equation}
\mathcal{F}(r_0) = \frac{2\varrho M(r_0 + 2M)}{r_0^2 + M^2 b}
+ \ln\frac{r_0 - M - M\varrho}{r_0 - M + M\varrho}\,,
\label{eq:F0}
\end{equation}
\begin{equation}
\mathcal{F}(r) = \frac{2\varrho M(r + 2M)}{r^2 + M^2 b}
+ \ln\frac{r - M - M\varrho}{r - M + M\varrho}\,.
\label{eq:Fr}
\end{equation}
In the interior region $\mathcal{F}(r_0)$ is a constant, so that $A_\varphi$ describes a uniform magnetic field inside the loop,
\begin{equation}
A_\varphi^{\rm int} = \frac{1}{2}B_*(r^2 + M^2 b)\sin^2\theta\,,
\label{eq:Aphi_int_B}
\end{equation}
with $B_*$ the uniform braneworld field strength~\cite{Turimov2018}.

\subsection{Effective potential for charged particles}

For a particle of mass $m$ and charge $q$ in the combined gravitational and electromagnetic field, the Hamilton--Jacobi equation reads~\cite{Babar2016,Turimov2020}
\begin{equation}
g^{\mu\nu}\left(\frac{\partial S}{\partial x^\mu} + qA_\mu\right)
\left(\frac{\partial S}{\partial x^\nu} + qA_\nu\right) + m^2 = 0\,.
\label{eq:HJ_charged}
\end{equation}
Since $A_\varphi$ is the only nonvanishing component, the energy $E=-p_t$ is conserved, while the canonical angular momentum becomes
$L=p_\varphi = mr^2\dot{\varphi}\sin^2\theta + qA_\varphi$. For equatorial motion, the radial equation takes the form
\begin{equation}
\dot{r}^2 = \mathcal{E}^2 - V_{\rm eff}(r)\,,
\label{eq:rdot_charged}
\end{equation}
with the charged-particle effective potential
\begin{equation}
V_{\rm eff}(r) = N_*^2(r)\left(1 + \frac{1}{r^2}
\left(\mathcal{L} + \frac{q}{m}A_\varphi\bigg|_{\theta=\pi/2}\right)^2\right)\,.
\label{eq:veff_charged}
\end{equation}
Substituting $\mu_*$ from Eq.~\eqref{eq:mu} and defining the magnetic coupling parameter
\begin{equation}
\omega = -\frac{3\pi I q}{8m}\, .
\label{eq:omega_def}
\end{equation}
The effective potential in the interior region ($r_+\leq r\leq r_0$) becomes
\begin{equation}
V_{\rm eff}^{\rm int}(r) = N_*^2(r)\left(1
+ \frac{\left(\mathcal{L} + \omega\,\mathcal{G}_{\rm int}(r)\right)^2}{r^2}\right)\,,
\label{eq:veff_int}
\end{equation}
where
\begin{equation}
\mathcal{G}_{\rm int}(r) = \frac{N_*(r_0)(r_0^2 + M^2 b)}{(1+b)^{3/2}\,M^3}
\,(r^2 + M^2 b)\,\mathcal{F}(r_0)\,.
\label{eq:Gint}
\end{equation}
Similarly, in the exterior region ($r\geq r_0$),
\begin{equation}
V_{\rm eff}^{\rm ext}(r) = N_*^2(r)\left(1
+ \frac{\left(\mathcal{L} + \omega\,\mathcal{G}_{\rm ext}(r)\right)^2}{r^2}\right)\,,
\label{eq:veff_ext}
\end{equation}
where, with the same prefactor as in Eq.~\eqref{eq:Gint} and only $\mathcal{F}(r_0)$ replaced by the running $\mathcal{F}(r)$,
\begin{equation}
\mathcal{G}_{\rm ext}(r) = \frac{N_*(r_0)(r_0^2 + M^2 b)}{(1+b)^{3/2}\,M^3}
\,(r^2 + M^2 b)\,\mathcal{F}(r)\,.
\label{eq:Gext}
\end{equation}
Both potentials reduce to the neutral case~\eqref{eq:veff_neutral} when $\omega=0$, vanish at the horizon, $V_{\rm eff}(r_+)=0$ for all $\omega$ and $\mathcal{L}$, and match continuously at $r=r_0$ since $\mathcal{G}_{\rm int}(r_0)=\mathcal{G}_{\rm ext}(r_0)$.

\begin{figure*}[t]
\centering
\includegraphics[width=0.45\linewidth]{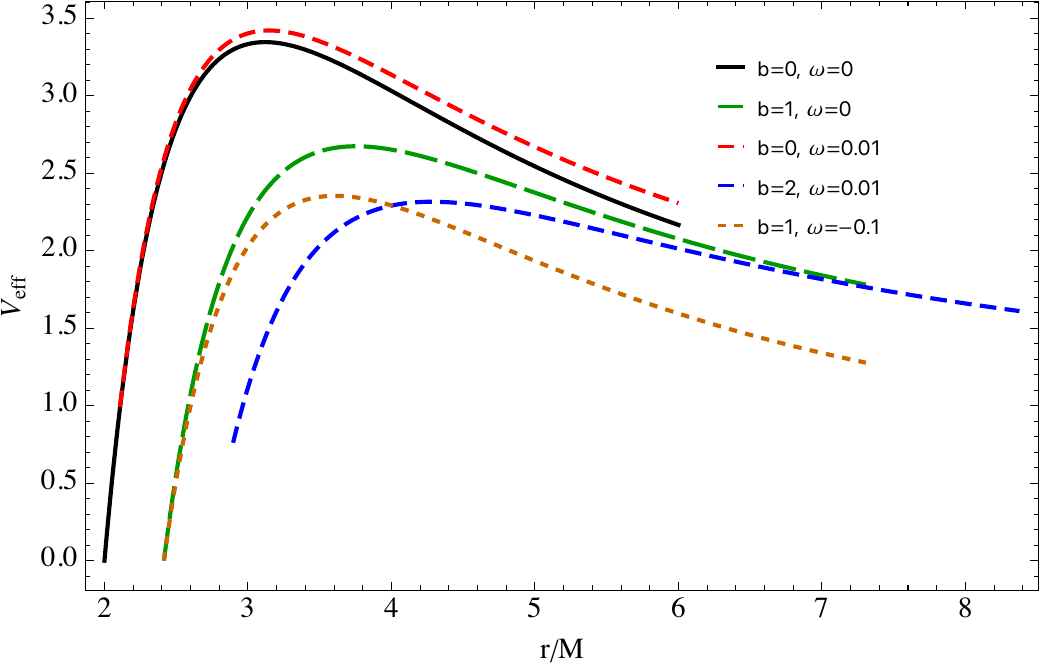}\hfill
\includegraphics[width=0.45\linewidth]{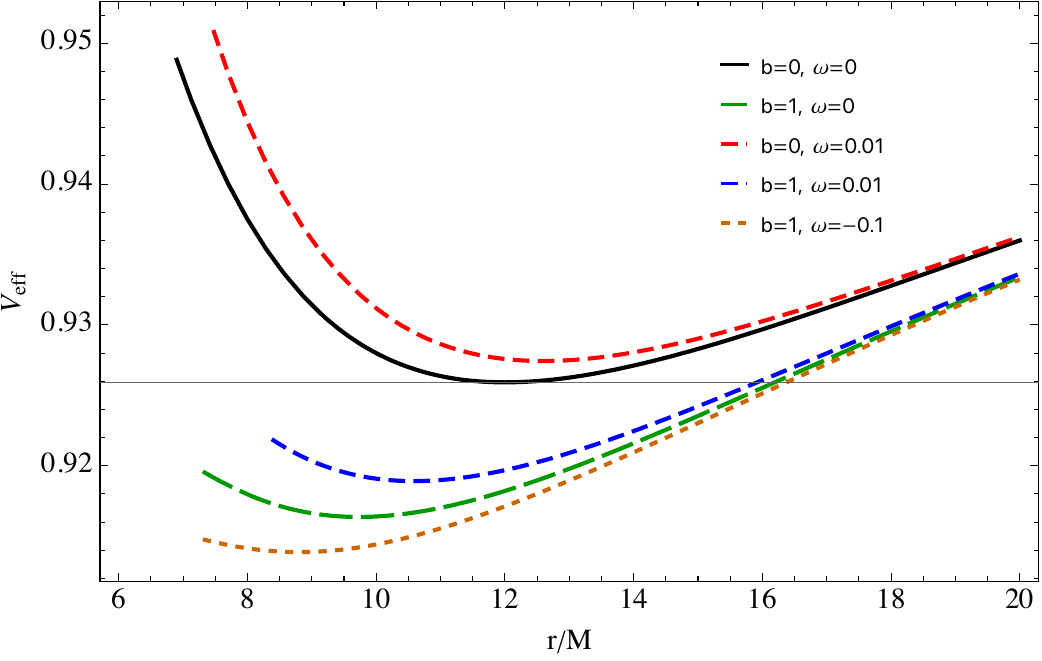}
\caption{The effective potential $V_{\rm eff}$ vs $r/M$ in the interior (left) and exterior
(right) regions of the current loop for selected values of $b$ and $\omega$ at
$\mathcal{L}=9M$.}
\label{fig:veff}
\end{figure*}

The radial dependence of $V_{\rm eff}$ is shown in Fig.~\ref{fig:veff} for selected $(b,\omega)$. Comparing the black ($b=0,\omega=0$) and green ($b=1,\omega=0$) curves isolates the pure braneworld effect: increasing $b$ lowers the peak of $V_{\rm eff}$ and shifts it outward.
Comparing black and red ($b=0,\omega=0.01$) isolates the magnetic coupling: a positive $\omega$ raises the barrier. The blue curve ($b=2,\omega=0.01$) combines both effects, while the orange dotted curve ($b=1,\omega=-0.1$) shows the braneworld effect together with a negative coupling, which suppresses the potential more strongly. In the interior region (left) increasing $b$ lowers and shifts the peak outward; a positive $\omega$ raises the barrier and a negative $\omega$ suppresses it; in the exterior region (right) the same trends persist near the ISCO minimum, where $b$ deepens and displaces the minimum outward while $\omega$ raises or lowers it according to its sign.

\begin{figure*}[t]
\centering
\includegraphics[width=0.45\linewidth]{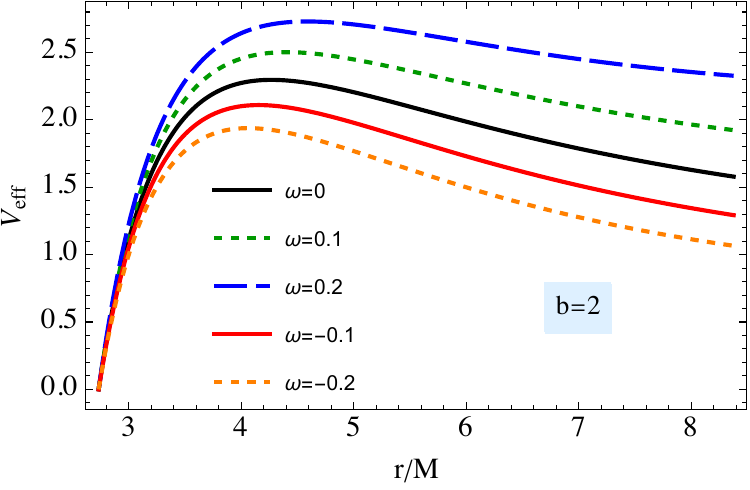}\hfill
\includegraphics[width=0.45\linewidth]{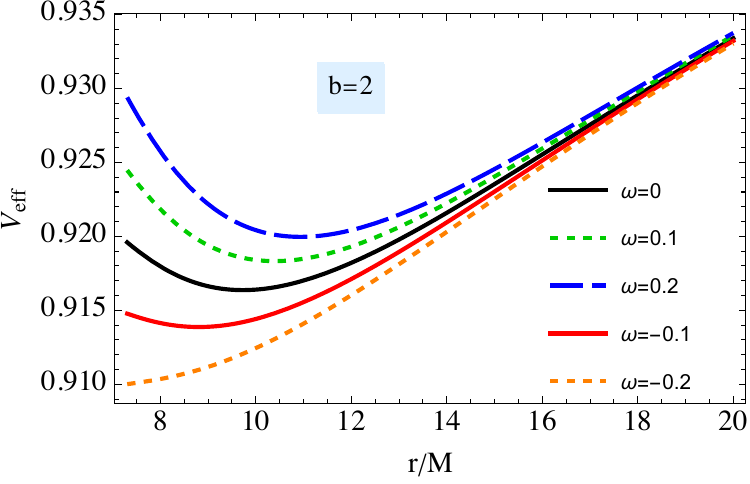}
\caption{The effective potential $V_{\rm eff}$ vs $r/M$ in the interior (left) and exterior (right) regions for fixed $b=2$ and selected $\omega$, with $\mathcal{L}=9M$.}
\label{fig:veff_omega}
\end{figure*}

The effect of $\omega$ alone is shown in Fig.~\ref{fig:veff_omega} for fixed $b=2$: in both regions a positive $\omega$ raises the potential and a negative $\omega$ suppresses it, with the neutral case in between; the effect is symmetric in sign but grows with $|\omega|$, consistent with the Lorentz interaction acting as an effective (para- or diamagnetic) correction to the centrifugal barrier~\cite{KhanChen2023,Haydarov2020}.

\section{Circular orbits and ISCO properties}
\label{sec:circular}

\subsection{Circular orbit conditions for charged particles}

For a charged particle in the equatorial plane, the circular orbit conditions are~\cite{Babar2016}
\begin{equation}
V_{\rm eff}(r) = \mathcal{E}^2\,, \qquad
\frac{dV_{\rm eff}}{dr}\bigg|_{r=r_c} = 0\,.
\label{eq:circ_conditions}
\end{equation}
The second condition implicitly fixes $\mathcal{L}$ at radius $r_c$. Substituting Eqs.~\eqref{eq:veff_int} and~\eqref{eq:veff_ext} one obtains
\begin{equation}
\frac{d}{dr}\left[N_*^2(r)\left(1 +
\frac{\bigl(\mathcal{L} + \omega\,\mathcal{G}(r)\bigr)^2}{r^2}\right)\right]_{r=r_c} = 0\,,
\label{eq:dVeff_zero}
\end{equation}
with $\mathcal{G}(r)$ standing for $\mathcal{G}_{\rm int}(r)$ or $\mathcal{G}_{\rm ext}(r)$ according to the region. For $\omega=0$ this reduces to the closed form~\eqref{eq:L_neutral}; for $\omega\neq0$ the logarithmic terms in $\mathcal{G}(r)$ make Eq.~\eqref{eq:dVeff_zero} transcendental, so $\mathcal{L}(r)$ is found numerically by root finding, and the energy follows from
\begin{equation}
\mathcal{E}(r_c) = \sqrt{V_{\rm eff}(r_c)}\,.
\label{eq:E_from_Veff}
\end{equation}
The ISCO radius follows by additionally imposing $d^2V_{\rm eff}/dr^2=0$, likewise solved numerically for $\omega\neq0$~\cite{Babar2016,Turimov2020}.

\begin{figure*}[t]
\centering
\includegraphics[width=0.45\linewidth]{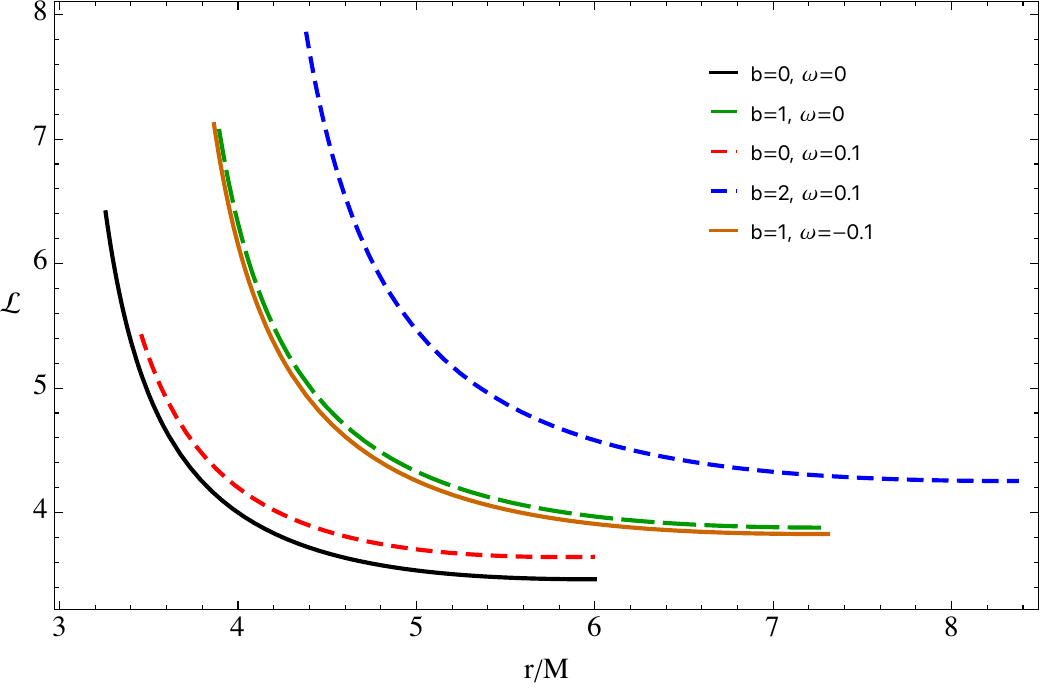}\hfill
\includegraphics[width=0.45\linewidth]{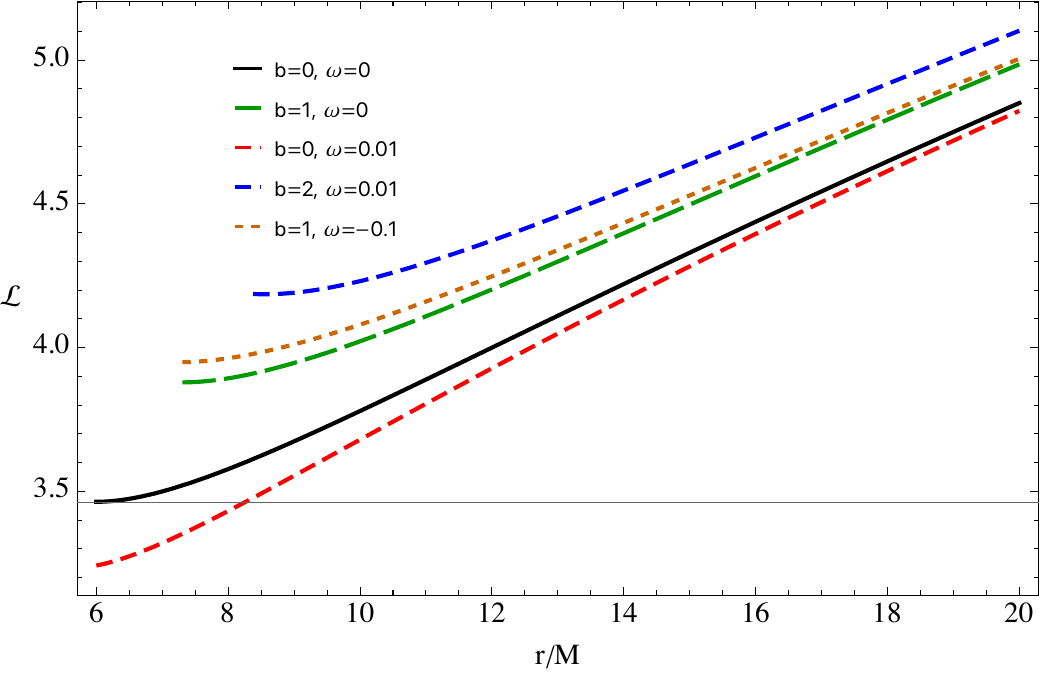}
\caption{Specific angular momentum $\mathcal{L}$ of charged particles on circular orbits vs $r/M$ in the interior (left) and exterior (right) regions, for selected values of $b$ and $\omega$.}
\label{fig:L_circ}
\end{figure*}

The specific angular momentum on circular orbits is shown in Fig.~\ref{fig:L_circ}. In the interior region (left) $\mathcal{L}$ decreases monotonically with $r$, diverging near the horizon and reaching its minimum at the ISCO; increasing $b$ shifts the whole profile to larger radii, a positive $\omega$ reduces $\mathcal{L}$, and a negative $\omega$ raises it. In the exterior region (right) $\mathcal{L}$ grows monotonically with $r$, as expected for Keplerian-like orbits, with the same qualitative dependence on $b$ and $\omega$. In both regions, the Schwarzschild neutral case ($b=0,\omega=0$) serves as the reference, and all deviations are monotonic in $b$ and $\omega$~\cite{Stuchlik2020}.

\begin{figure}[t]
\centering
\includegraphics[width=0.9\linewidth]{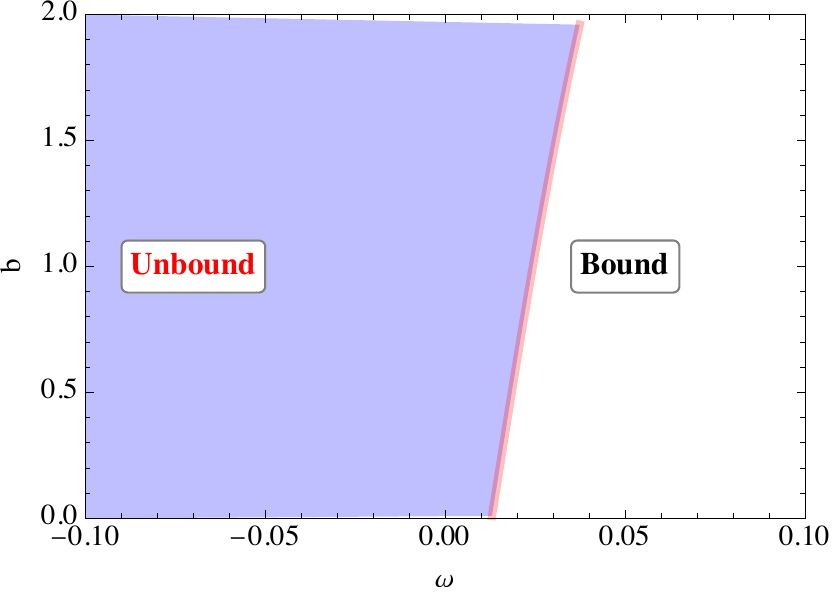}
\caption{The boundary $\mathcal{E}=1$ in the $(b,\omega)$ plane for the interior region at fixed loop radius. The shaded region denotes unbound circular-orbit energy ($\mathcal{E}>1$); the white region corresponds to bound orbits ($\mathcal{E}<1$).}
\label{fig:bound_in}
\end{figure}

The boundary $\mathcal{E}(b,\omega)=1$ in the $(b,\omega)$ plane for the interior region, evaluated at the fixed loop location $r_0$, is shown in Fig.~\ref{fig:bound_in}. In the shaded region $\mathcal{E}>1$, the electromagnetic interaction transfers enough energy to render the orbit unbound, so that the accretion efficiency $\eta=1-\mathcal{E}$ is negative~\cite{Bambi2014}. The white region ($\mathcal{E}<1$) corresponds to physically relevant bound orbits. The boundary shifts to the right with increasing $b$, so a larger tidal charge requires a more positive $\omega$ to keep the fixed-radius orbit bound.

\begin{figure*}[t]
\centering
\includegraphics[width=0.45\linewidth]{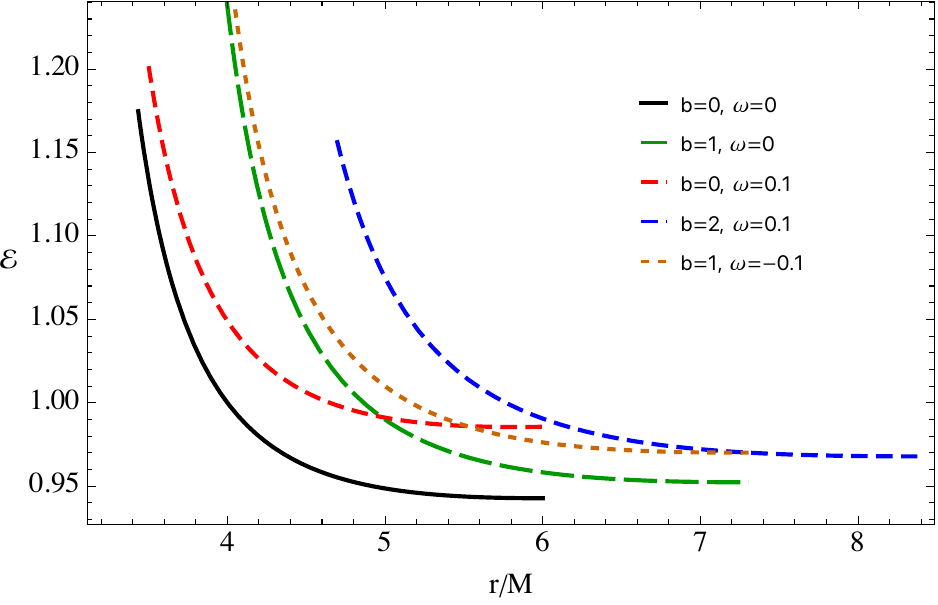}\hfill
\includegraphics[width=0.45\linewidth]{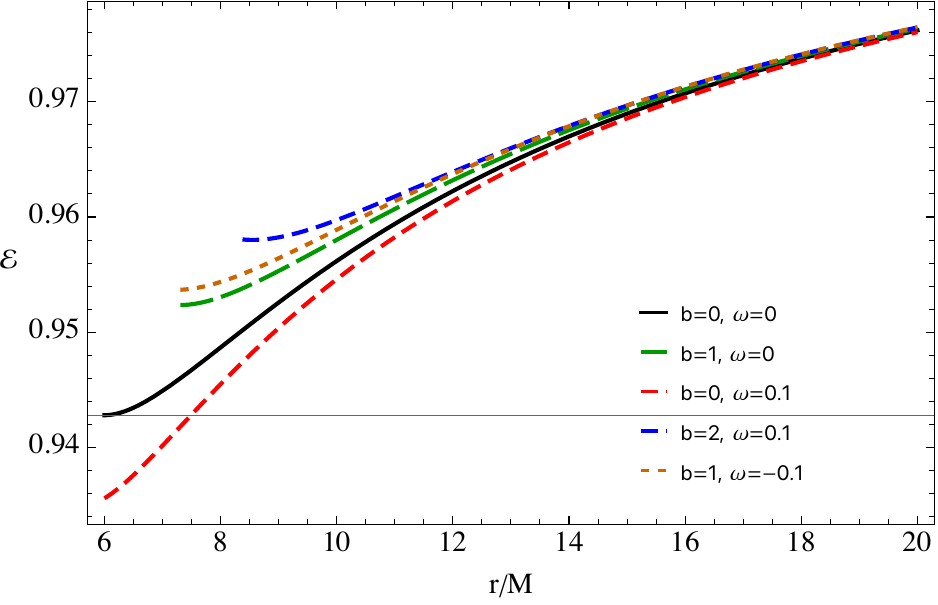}
\caption{Specific energy $\mathcal{E}$ of charged particles on circular orbits vs $r/M$ in the interior (left) and exterior (right) regions, for selected $b$ and $\omega$. The horizontal line at $\mathcal{E}=1$ marks the bound--unbound threshold.}
\label{fig:E_circ}
\end{figure*}

The specific energy on circular orbits is shown in Fig.~\ref{fig:E_circ}. In the interior region (left) all curves decrease with $r$ and exceed unity near the horizon; the neutral Schwarzschild curve reaches $\mathcal{E}<1$ at larger radii, and increasing $b$ shifts this crossing outward. A positive $\omega$ raises the energy profile while a negative $\omega$ lowers it, and for large positive $\omega$ combined with large $b$ the energy remains above unity across the whole interior region, consistent with the unbound zone of Fig.~\ref{fig:bound_in}. In the exterior region (right) all curves lie below unity and rise toward $\mathcal{E}\to1$ at large $r$. The ISCO energy---the leftmost point of each curve---\emph{increases} with $b$ at fixed $\omega$ (e.g.\ $\mathcal{E}_{\rm ISCO}=0.943,0.952,0.958$ for $b=0,1,2$ at $\omega=0$); equivalently the accretion efficiency $\eta=1-\mathcal{E}_{\rm ISCO}$ \emph{decreases} with $b$, so that the negative tidal charge---which displaces the ISCO outward---lowers the binding energy of the innermost orbit relative to Schwarzschild.

\subsection{ISCO radius dependence on $\omega$}

The charged-particle ISCO is fixed by the three simultaneous conditions
\begin{equation}
V_{\rm eff}(r) = \mathcal{E}^2\,, \quad
\frac{dV_{\rm eff}}{dr} = 0\,, \quad
\frac{d^2V_{\rm eff}}{dr^2} = 0\,,
\label{eq:isco_three}
\end{equation}
solved self-consistently for $(r_{\rm ISCO},\mathcal{L}_{\rm ISCO},\mathcal{E}_{\rm ISCO})$ as
functions of $b$ and $\omega$. For $\omega=0$ the system reduces to the closed form~\eqref{eq:risco}; for $\omega\neq0$ the logarithmic terms make it transcendental, and it is solved numerically for each $(b,\omega)$~\cite{Babar2016,Turimov2020}. In practice, for each $(b,\omega)$ we first solve $dV_{\rm eff}/dr=0$ for $\mathcal{L}$ at a trial radius and then locate the root of $d^2V_{\rm eff}/dr^2=0$, evaluating $\mathcal{L}_{\rm ISCO}$ and $\mathcal{E}_{\rm ISCO}$ there. We emphasize an exact consistency check that any numerical implementation must satisfy: since the coupling term $\omega\,\mathcal{G}$ vanishes at $\omega=0$, the interior and exterior ISCO curves must \emph{coincide} at $\omega=0$ on the neutral value $r_0=r_{\rm ISCO}(b)$ of Eq.~\eqref{eq:risco}, and the exterior ISCO can never fall below the loop radius $r_0$.

\begin{figure}[t]
\centering
\includegraphics[width=0.9\linewidth]{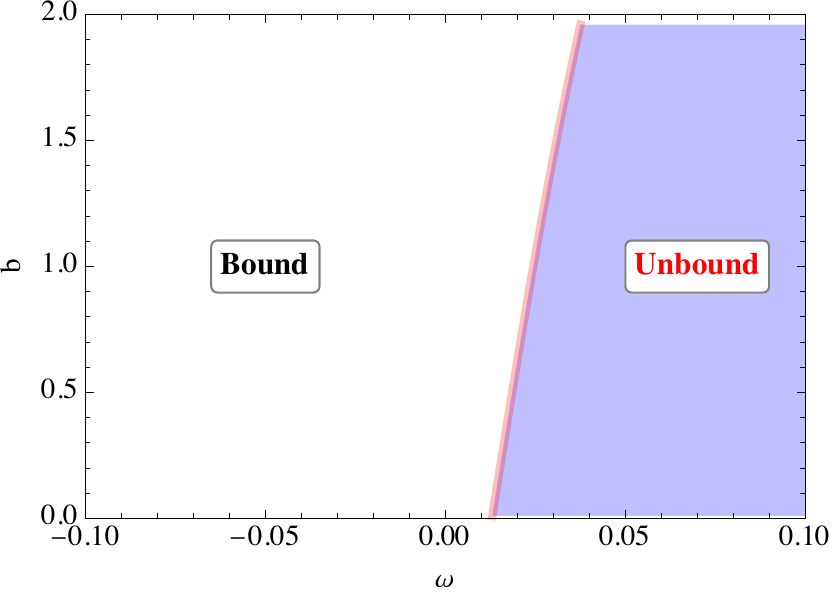}
\caption{The boundary $\mathcal{E}_{\rm ISCO}=1$ in the $(b,\omega)$ plane, evaluated at the true ISCO radius $r_{\rm ISCO}(b,\omega)$. The shaded region corresponds to unbound ISCO orbits and the white region to bound orbits.}
\label{fig:bound_eisco}
\end{figure}

The boundary $\mathcal{E}_{\rm ISCO}(b,\omega)=1$ in the $(b,\omega)$ plane, evaluated at the true ISCO radius $r_{\rm ISCO}(b,\omega)$, is shown in Fig.~\ref{fig:bound_eisco}. The shaded region corresponds to unbound ISCO orbits ($\mathcal{E}_{\rm ISCO}>1$) and the white region to bound orbits. The boundary shifts to larger $\omega$ as $b$ increases, so a stronger tidal charge stabilizes the ISCO and requires a larger positive coupling to unbind it. This boundary differs from the fixed-radius boundary of Fig.~\ref{fig:bound_in} because here the loop location is held at $r_0$ while the ISCO itself moves with $\omega$; the two therefore respond oppositely, showing that the unbinding of the ISCO is a combined effect of the direct electromagnetic interaction and the $\omega$-induced shift of the ISCO location~\cite{Bambi2014,KhanChen2023}.

The sign of $\omega$ is set by the sign of the charge $q$: from Eq.~\eqref{eq:omega_def}, a positively charged particle ($q>0$) with a positive current gives $\omega<0$, corresponding to the bound region of Fig.~\ref{fig:bound_eisco}, whereas a negatively charged particle ($q<0$) gives $\omega>0$ and, for sufficiently large $|\omega|$, an unbound orbit. Bound ISCO orbits therefore favor positively charged particles in the loop field, with implications for the charge composition of accreting plasma in the braneworld spacetime~\cite{KhanChen2023,Rayimbaev2021}.

\begin{figure*}[t]
\centering
\includegraphics[width=0.45\linewidth]{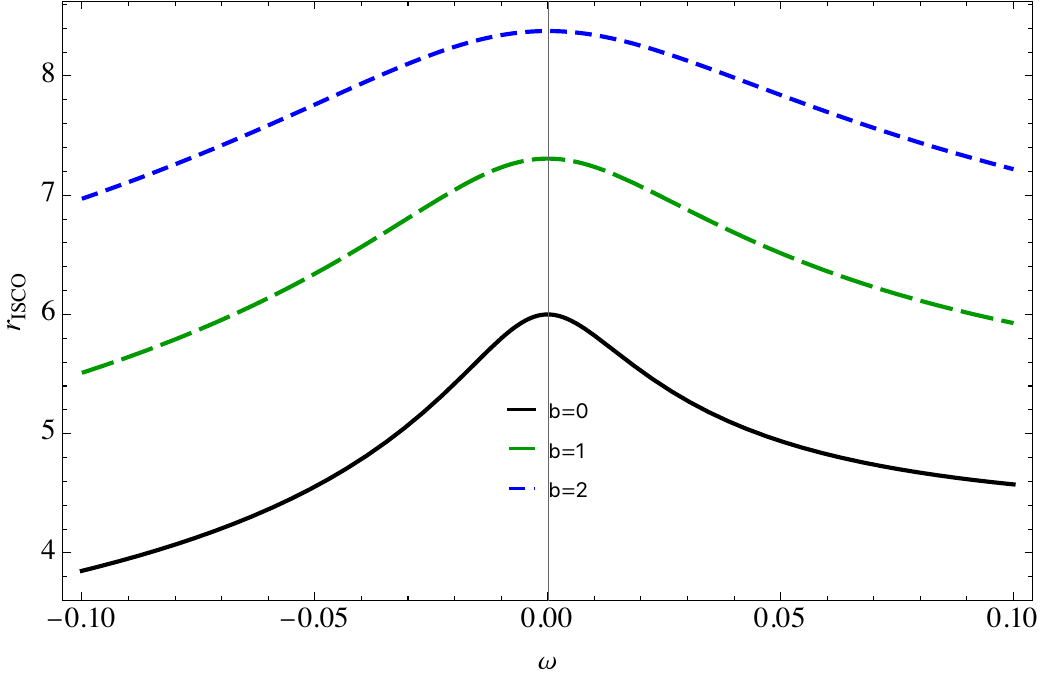}\hfill
\includegraphics[width=0.45\linewidth]{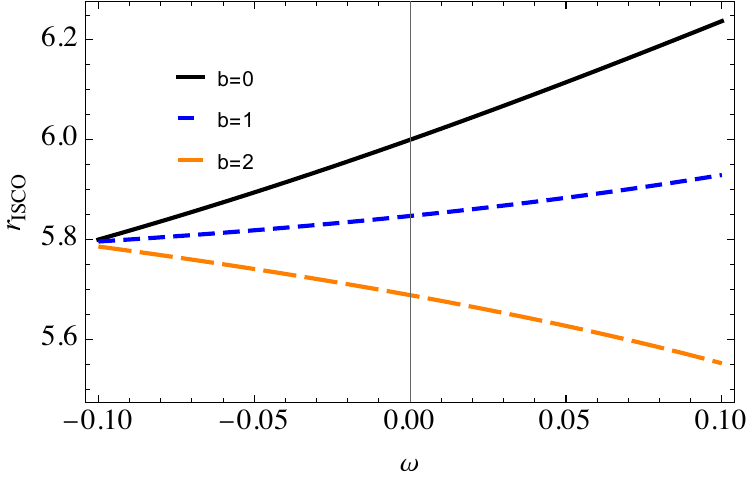}
\caption{The ISCO radius $r_{\rm ISCO}$ as a function of $\omega$ for selected values of $b$, in the interior (left) and exterior (right) regions. At $\omega=0$ each curve recovers the neutral ISCO radius of Eq.~\eqref{eq:risco}.}
\label{fig:risco_omega}
\end{figure*}

The dependence of $r_{\rm ISCO}$ on $\omega$ is shown in Fig.~\ref{fig:risco_omega}. In the interior region (left) each curve peaks near $\omega\approx0$ and decreases on both sides: both signs of the coupling pull the interior ISCO inward, more strongly for larger $|\omega|$, while the peak value increases with $b$. At $\omega=0$ each curve recovers the neutral value of Eq.~\eqref{eq:risco}. In the exterior region (right) the curves start from the same neutral value at $\omega=0$ and, constrained to $r\geq r_0$, respond to the coupling through the running function $\mathcal{G}_{\rm ext}(r)$; the tidal charge and the magnetic coupling compete, so that the ordering with $b$ can differ from the interior case.

\subsection{ISCO angular momentum}

\begin{figure*}[t]
\centering
\includegraphics[width=0.45\linewidth]{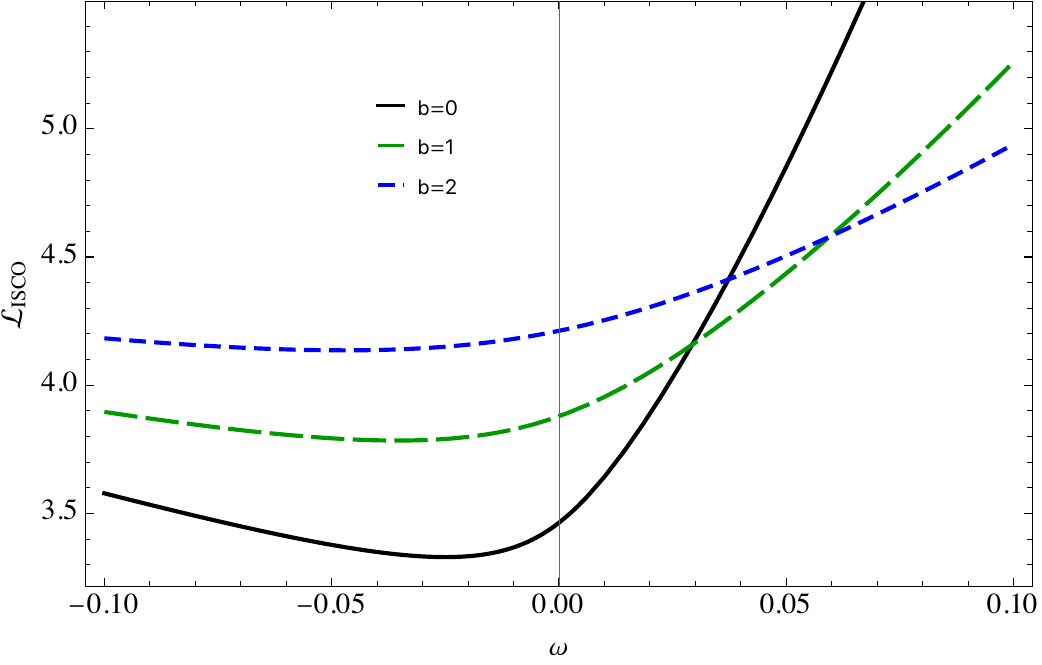}\hfill
\includegraphics[width=0.45\linewidth]{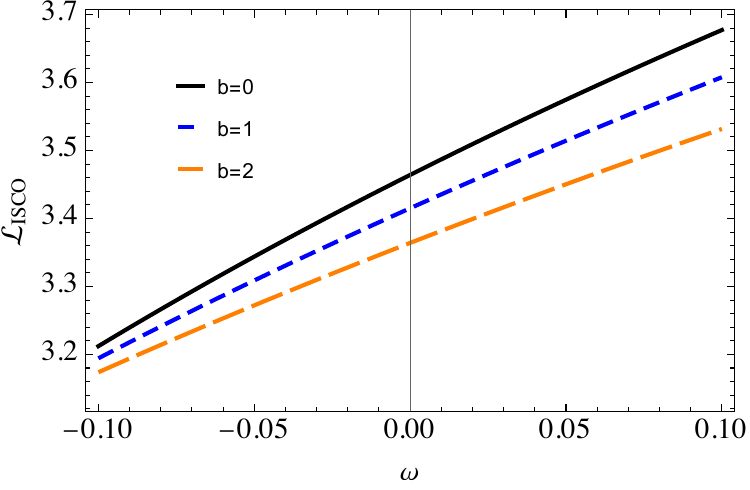}
\caption{The ISCO angular momentum $\mathcal{L}_{\rm ISCO}$ as a function of $\omega$ for selected values of $b$, in the interior (left) and exterior (right) regions.}
\label{fig:lisco_omega}
\end{figure*}

The specific angular momentum at the ISCO is obtained from the same procedure. For $\omega=0$ it reduces to Eq.~\eqref{eq:L_neutral} at $r=r_{\rm ISCO}(b)$; for $\omega\neq0$ the coupling modifies both the ISCO location and the angular momentum required there. The dependence of $\mathcal{L}_{\rm ISCO}$ on $\omega$ is shown in Fig.~\ref{fig:lisco_omega}. In the interior region (left) each curve has a minimum near $\omega\approx0$ and rises for positive $\omega$, producing an asymmetric U-shaped profile; at fixed $\omega$, larger $b$ corresponds to larger $\mathcal{L}_{\rm ISCO}$, consistent with the outward ISCO shift. The exterior curves again originate from the same neutral value at $\omega=0$ and vary nearly linearly with $\omega$, reflecting the approximately additive electromagnetic correction in that regime~\cite{Stuchlik2020}.

\subsection{ISCO energy}

\begin{figure*}[t]
\centering
\includegraphics[width=0.45\linewidth]{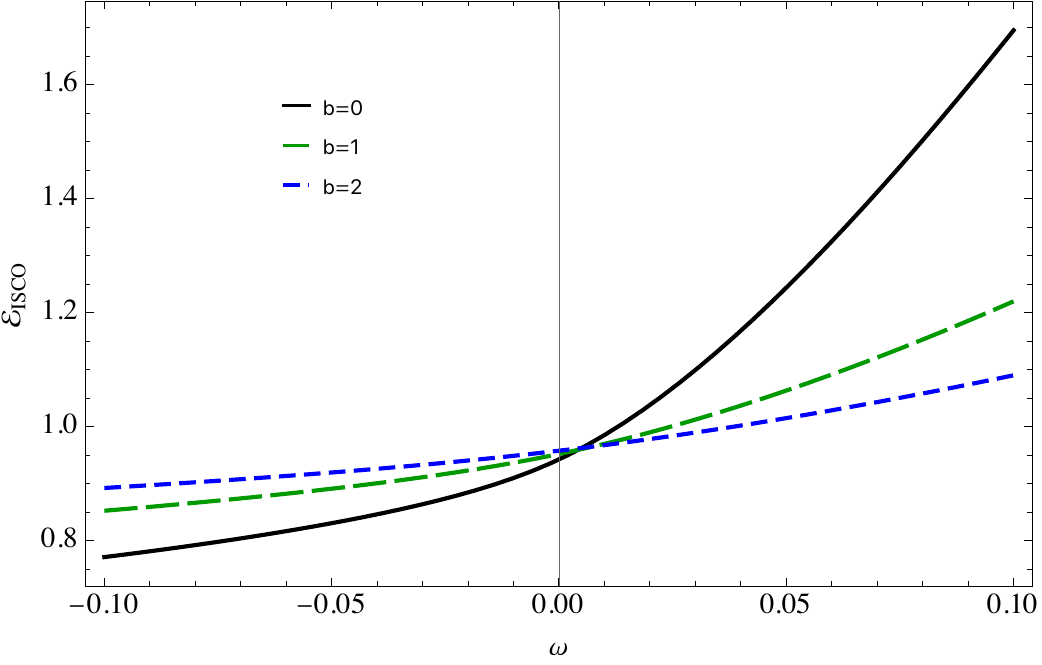}\hfill
\includegraphics[width=0.45\linewidth]{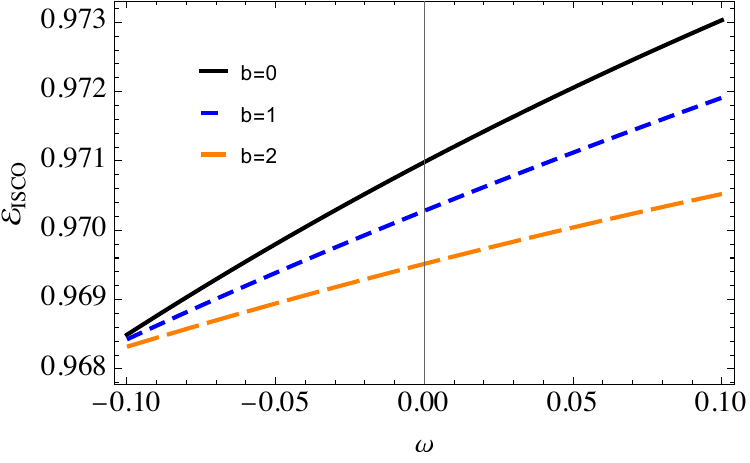}
\caption{The ISCO energy $\mathcal{E}_{\rm ISCO}$ as a function of $\omega$ for selected $b$, in the interior (left) and exterior (right) regions. The horizontal line in the left panel marks the bound--unbound threshold $\mathcal{E}_{\rm ISCO}=1$.}
\label{fig:eisco_omega}
\end{figure*}

The specific energy at the ISCO, $\mathcal{E}_{\rm ISCO}=\sqrt{V_{\rm eff}(r_{\rm ISCO})}$, follows self-consistently; for $\omega=0$ it reduces to Eq.~\eqref{eq:E_neutral} at $r=r_{\rm ISCO}(b)$, giving $\mathcal{E}_{\rm ISCO}=0.943,0.952,0.958$ for $b=0,1,2$. The accretion efficiency is
\begin{equation}
\eta = 1 - \mathcal{E}_{\rm ISCO}\,,
\label{eq:efficiency}
\end{equation}
positive only for bound ISCO orbits, $\mathcal{E}_{\rm ISCO}<1$~\cite{Bambi2014}. The dependence of $\mathcal{E}_{\rm ISCO}$ on $\omega$ is shown in Fig.~\ref{fig:eisco_omega}. In the interior region (left) all curves increase monotonically with $\omega$ and cross the threshold $\mathcal{E}_{\rm ISCO}=1$ at a critical $\omega_{\rm cr}(b)$ near $\omega\approx0$; for $\omega>\omega_{\rm cr}$, the orbit is unbound. All curves converge near $\omega\approx0$ on the neutral values, consistent with Fig.~\ref{fig:bound_eisco}. In the exterior region (right) all curves lie below unity, confirming that exterior ISCO orbits remain bound over the range considered, with a weaker $\omega$-dependence reflecting the more gradual radial profile of $\mathcal{G}_{\rm ext}(r)$.

\subsection{ISCO orbital frequency}

\begin{figure*}[t]
\centering
\includegraphics[width=0.45\linewidth]{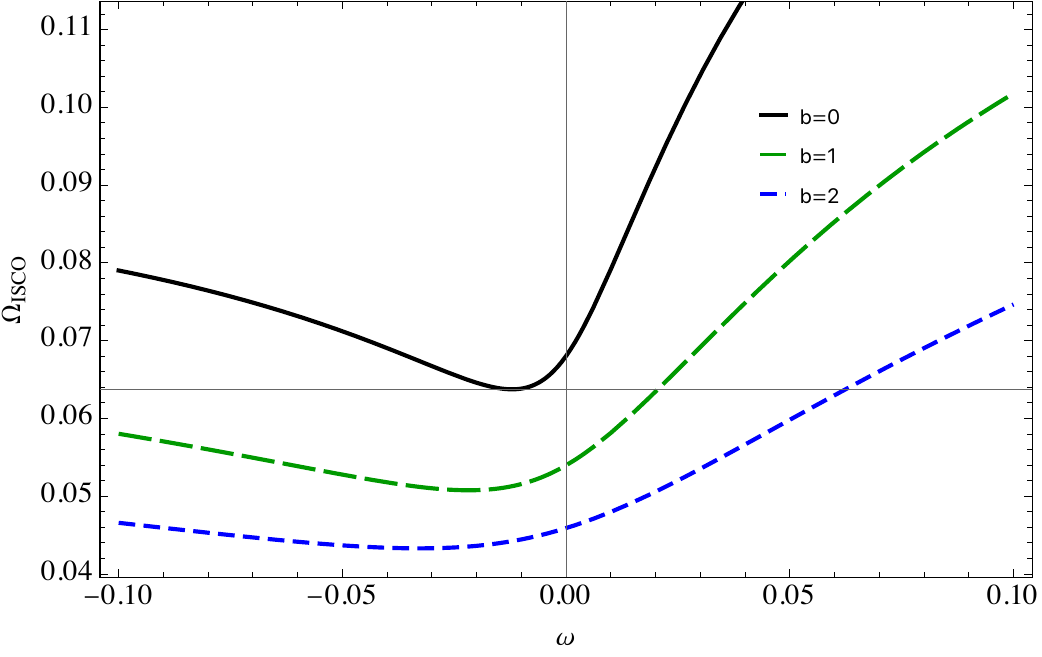}\hfill
\includegraphics[width=0.45\linewidth]{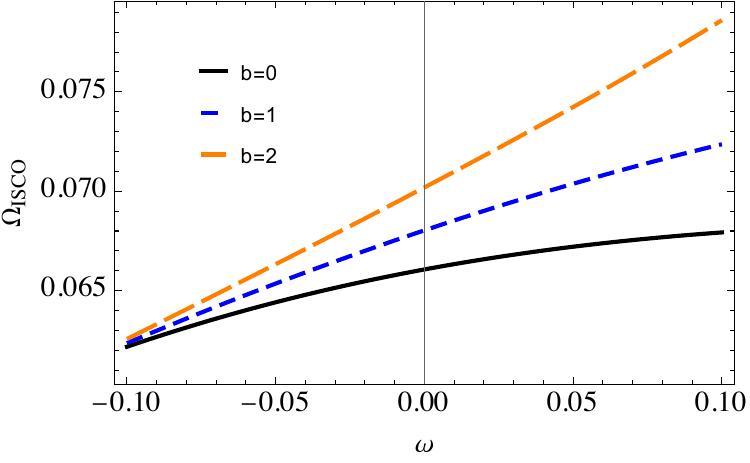}
\caption{The ISCO orbital frequency $\Omega_{\rm ISCO}$ as a function of $\omega$ for selected $b$, in the interior (left) and exterior (right) regions. At $\omega=0$ the Schwarzschild value $\Omega_{\rm ISCO}=6^{-3/2}M^{-1}$ is recovered for $b=0$.}
\label{fig:omegaisco}
\end{figure*}

The angular velocity of a charged particle on a circular orbit, as measured by a distant observer, is
\begin{equation}
\Omega_\varphi = \frac{d\varphi}{dt} =
\frac{N_*^2(r)\left(\mathcal{L} + \frac{q}{m}A_\varphi\right)}{\mathcal{E}\,r^2}\,,
\label{eq:Omega_phi}
\end{equation}
evaluated at $r=r_{\rm ISCO}(b,\omega)$ with the self-consistent $\mathcal{L}_{\rm ISCO}$ and $\mathcal{E}_{\rm ISCO}$. For $\omega=0$, $\Omega_{\rm ISCO}$ depends on $b$ only through the ISCO radius and decreases with $b$, since a larger $b$ pushes the ISCO outward where the frequency is lower; at $b=0$ the Schwarzschild value $\Omega_{\rm ISCO}=6^{-3/2}M^{-1}\simeq0.068\,M^{-1}$ is recovered. The dependence on $\omega$ is shown in Fig.~\ref{fig:omegaisco}. In the interior region (left) each curve has a minimum near $\omega\approx0$ and rises steeply for positive $\omega$, mirroring $r_{\rm ISCO}$ and $\mathcal{L}_{\rm ISCO}$; the ordering $\Omega_{\rm ISCO}(b=0)>\Omega_{\rm ISCO}(b=1)>\Omega_{\rm ISCO}(b=2)$ is recovered at $\omega=0$. In the exterior region (right) the curves again start from the same neutral values at $\omega=0$ and are governed by the competition between the tidal charge and the magnetic coupling.

\section{Fundamental frequencies}
\label{sec:frequencies}

Having established the circular orbits and ISCO, we now compute the three fundamental frequencies governing test-particle motion in the current-loop field: the azimuthal (Keplerian) frequency $\Omega_\phi$ and the radial and vertical epicyclic frequencies $\Omega_r$ and $\Omega_\theta$. These are the building blocks of the twin-peak QPO models of Sec.~\ref{sec:qpo}~\cite{Stella1999,AbramowiczKluzniak2001,Torok2005}.

\subsection{Keplerian frequency}
\label{subsec:keplerian}

\begin{figure*}[t]
\centering
\includegraphics[width=0.9\linewidth]{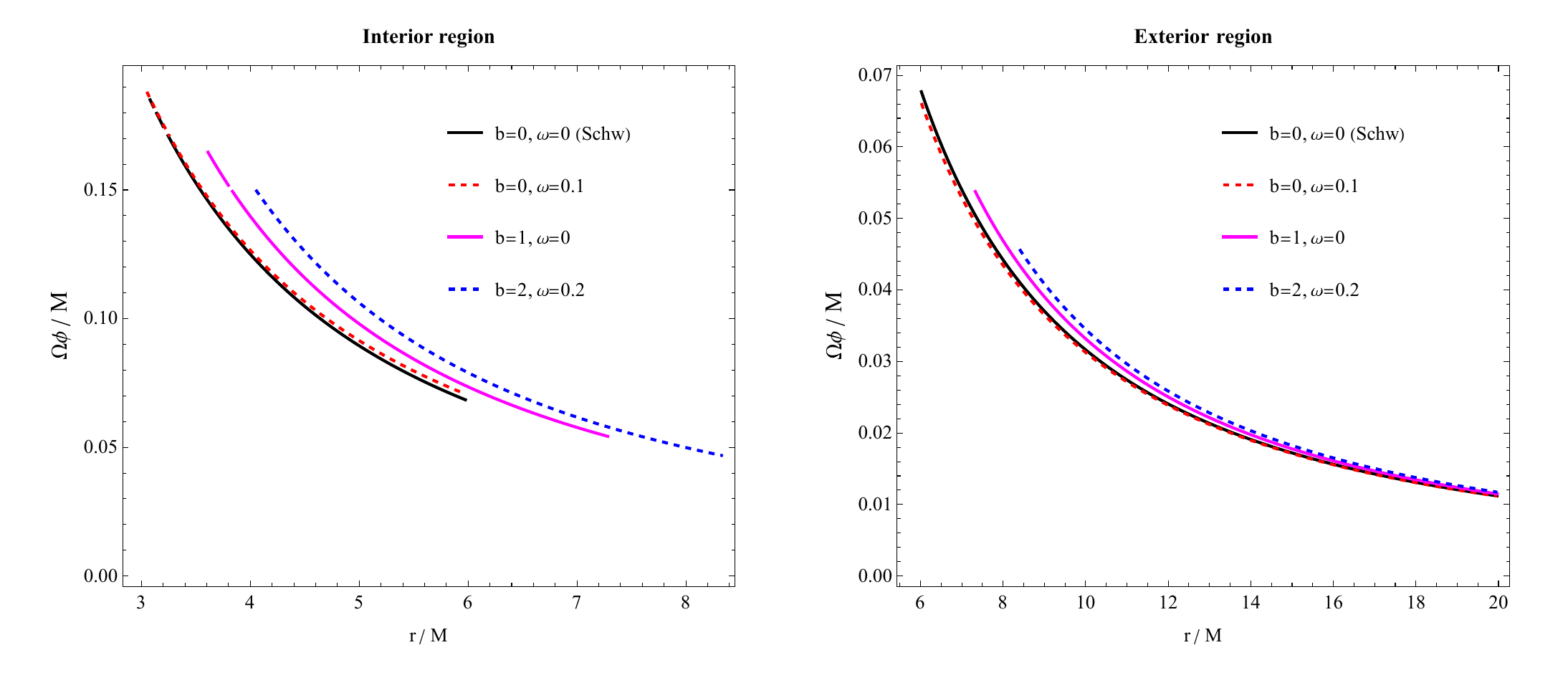}
\caption{Radial profile of the orbital frequency $\Omega_\phi/M$ for a charged particle in the interior (left) and exterior (right) regions of the current loop, for selected values of $b$ and $\omega$. The tidal charge raises the frequency at fixed radius, while the magnetic coupling produces only a slight modification.}
\label{fig:omega_phi}
\end{figure*}

For a static, spherically symmetric spacetime the orbital angular velocity is fixed by the radial gradients of the metric coefficients,
\begin{equation}
\Omega_\phi = \sqrt{\frac{-\,\partial_r g_{tt}}{\partial_r g_{\varphi\varphi}}} = \sqrt{\frac{1}{2r}\,\frac{dN_*^2}{dr}}\,.
\label{eq:omega_phi_general}
\end{equation}
Substituting the tidal charged lapse, the neutral frequency takes the closed form
\begin{equation}
\Omega_\phi = \sqrt{\frac{M}{r^{3}} + \frac{M^{2}b}{r^{4}}}\,,
\label{eq:omega_phi_neutral}
\end{equation}
which reduces to $\Omega_\phi=\sqrt{M/r^3}$ as $b\to0$. In contrast to the Reissner--Nordstr\"om case, where the charge term enters with a negative sign and lowers the frequency, here the tidal charge appears with a strictly positive coefficient and \emph{raises} $\Omega_\phi$ at fixed radius---the frequency-domain signature of the stronger gravitational field produced by $Q_*<0$~\cite{Dadhich2000}. When the particle is charged, the frequency acquires the loop contribution through $\mathcal{L}+\omega\,\mathcal{G}(r)$,
\begin{equation}
\Omega_\phi = \frac{N_*^{2}(r)\bigl(\mathcal{L} + \omega\,\mathcal{G}(r)\bigr)}
{\mathcal{E}\,r^{2}}\,,
\label{eq:omega_phi_charged}
\end{equation}
reducing to Eq.~\eqref{eq:omega_phi_neutral} at $\omega=0$. In physical units,
\begin{equation}
\nu_\phi = \frac{c^{3}}{2\pi GM}\,\Omega_\phi\,.
\label{eq:hz_phi}
\end{equation}
The radial profile of $\Omega_\phi/M$ is shown in Fig.~\ref{fig:omega_phi}. In both regions the frequency decreases with $r$; the $b=1,2$ curves lie systematically above Schwarzschild in accordance with Eq.~\eqref{eq:omega_phi_neutral}, whereas switching on $\omega$ alone leaves the profile almost unchanged, so the braneworld geometry dominates the bare orbital frequency.

\subsection{Radial and vertical epicyclic frequencies}
\label{subsec:epicyclic}

Perturbing a stable circular orbit, $r\to r_0+\delta r$ and $\theta\to\pi/2+\delta\theta$, and expanding the effective potential to second order,
\begin{align}
V_{\rm eff}(r,\theta) &= V_{\rm eff}(r_0,\tfrac{\pi}{2})
+\tfrac{1}{2}\delta r^{2}\,\partial_r^{2}V_{\rm eff}\big|_{r_0,\pi/2}
\nonumber\\
&\quad+\tfrac{1}{2}\delta\theta^{2}\,\partial_\theta^{2}V_{\rm eff}\big|_{r_0,\pi/2}
+\mathcal{O}(\delta^{3})\,,
\label{eq:taylor}
\end{align}
the displacements obey decoupled oscillator equations whose frequencies, measured by a distant observer, are~\cite{Stella1999,Stuchlik2020}
\begin{equation}
\Omega_r^{2} = \frac{N_*^{4}(r)}{2\,\mathcal{E}^{2}}\,
\partial_r^{2}V_{\rm eff}\Big|_{\theta=\pi/2}\,, \quad
\Omega_\theta^{2} = \frac{N_*^{2}(r)}{2\,\mathcal{E}^{2}r^{2}}\,
\partial_\theta^{2}V_{\rm eff}\Big|_{\theta=\pi/2}\,.
\label{eq:epicyclic_def}
\end{equation}
For a neutral particle, the derivatives are closed-form,
\begin{align}
\Omega_r^{2} &= \frac{M}{r^{3}} - \frac{6M^{2}}{r^{4}}
- \frac{9M^{3}b}{r^{5}} - \frac{4M^{4}b^{2}}{r^{6}}
\nonumber\\
&= \frac{M}{r^{6}}\Bigl(r^{3} - 6Mr^{2} - 9M^{2}b\,r - 4M^{3}b^{2}\Bigr)\,,
\label{eq:omega_r}
\end{align}
while the vertical frequency coincides with the azimuthal one,
\begin{equation}
\Omega_\theta^{2} = \Omega_\phi^{2} = \frac{M}{r^{3}} + \frac{M^{2}b}{r^{4}}
\qquad(\omega=0)\,,
\label{eq:omega_theta_neutral}
\end{equation}
so that the nodal frequency $\Omega_\phi-\Omega_\theta$ vanishes for neutral geodesics. Equation~\eqref{eq:omega_r} provides an internal check: $\Omega_r^{2}=0$ reproduces the neutral ISCO of Eqs.~\eqref{eq:risco}--\eqref{eq:eta}, giving $r_{\rm ISCO}=6M$ at $b=0$ and $r_{\rm ISCO}\simeq7.31M$ at $b=1$, in agreement with Fig.~\ref{fig:isco_b}.

Once the particle couples to the loop, the $\sin^2\theta$ dependence of $A_\varphi$ prevents the latitudinal derivative from collapsing onto the azimuthal value, and the vertical frequency acquires the closed form
\begin{equation}
\Omega_\theta^{2} = \frac{N_*^{4}(r)}{\mathcal{E}^{2}r^{4}}
\bigl(\mathcal{L} + \omega\,\mathcal{G}\bigr)\bigl(\mathcal{L} - \omega\,\mathcal{G}\bigr)
= \Omega_\phi^{2}\,\frac{\mathcal{L} - \omega\,\mathcal{G}}
{\mathcal{L} + \omega\,\mathcal{G}}\,.
\label{eq:omega_theta}
\end{equation}
The splitting between $\Omega_\theta$ and $\Omega_\phi$ is thus controlled entirely by $\omega\,\mathcal{G}$: a nonzero nodal frequency is generated purely by the electromagnetic sector, even though the background is static and non-rotating. This is a genuinely new feature relative to the neutral geodesic case---it is the current-loop field, not any rotation of the spacetime, that revives the nodal precession~\cite{Stella1999,Tursunov2018}. The radial frequency for a charged particle, which does not admit a compact closed form, is evaluated numerically at the circular-orbit radius using the self-consistent $\mathcal{L}$ and $\mathcal{E}$ of Sec.~\ref{sec:circular}. The epicyclic frequencies are expressed in physical units through
\begin{equation}
\nu_i = \frac{c^{3}}{2\pi GM}\,\Omega_i\,, \qquad i=r,\theta\,.
\label{eq:hz_epicyclic}
\end{equation}
Equations~\eqref{eq:omega_phi_charged}, \eqref{eq:omega_r}--\eqref{eq:omega_theta}, together
with the numerical radial frequency, furnish the complete set of fundamental frequencies for the QPO models below.

\section{Quasi-periodic oscillation models}
\label{sec:qpo}

Twin-peak high-frequency QPOs manifest as two commensurable peaks, an upper frequency $\nu_U$ and a lower frequency $\nu_L$, whose ratio frequently locks near $3\!:\!2$~\cite{AbramowiczKluzniak2001,Torok2005,Belloni2012}. Within the geodesic framework, these peaks combine the fundamental frequencies of Sec.~\ref{sec:frequencies}, with each combination corresponding to a distinct excitation scenario in the accretion flow~\cite{Stella1999,IngramMotta2019}. Below we collect the predictions of five established resonance models for the tidal charged braneworld black hole threaded by the current-loop field, using the Schwarzschild case $(b=0,\omega=0)$ as the baseline.

\paragraph*{Relativistic precession (RP) model.}
The RP model attributes twin-peak QPOs to the orbital dynamics of localized inhomogeneities on mildly eccentric and inclined geodesics~\cite{Stella1998,Stella1999,Motta2014}. The eccentricity drives periastron precession, and the inclination drives nodal precession; superposed on the fast azimuthal revolution, these generate
\begin{equation}
\nu_U = \nu_\phi\,, \qquad \nu_L = \nu_\phi - \nu_r\,,
\label{eq:rp}
\end{equation}
with the low-frequency modulation associated with the nodal frequency $\nu_\phi-\nu_\theta$. In a static, neutral spacetime, this nodal frequency vanishes identically; a distinctive feature of the present configuration is that the dipolar loop field revives it through the coupling $\omega\,\mathcal{G}$ [Eq.~\eqref{eq:omega_theta}], providing a purely geometric, resonance-free account of QPOs~\cite{Stella1999}.

\paragraph*{Epicyclic-resonance (ER) models.}
The ER models assume a geometrically thick disk and attribute the twin peaks to a nonlinear resonance between the radial and vertical epicyclic modes, which lock into a low-order commensurability at particular resonant radii~\cite{AbramowiczKluzniak2001,Torok2005}. Three commonly studied variants are
\begin{align}
\text{ER2:}\quad & \nu_U = 2\nu_\theta - \nu_r\,, & \nu_L &= \nu_r\,, \label{eq:er2}\\
\text{ER3:}\quad & \nu_U = \nu_\theta + \nu_r\,,  & \nu_L &= \nu_\theta\,, \label{eq:er3}\\
\text{ER4:}\quad & \nu_U = \nu_\theta + \nu_r\,,  & \nu_L &= \nu_\theta - \nu_r\,. \label{eq:er4}
\end{align}
Since these involve $\nu_\theta$ explicitly, they are directly sensitive to the magnetic coupling through the vertical frequency of Eq.~\eqref{eq:omega_theta}; in the neutral limit $\nu_\theta\to\nu_\phi$ they reduce to combinations of the orbital and radial frequencies.

\paragraph*{Warped-disc (WD) model.}
The WD model interprets high-frequency QPOs as nonlinear resonant coupling between a relativistically warped disc and the intrinsic oscillation modes of the disk~\cite{IngramMotta2019}, with the upper and lower peaks
\begin{equation}
\nu_U = 2\nu_\phi - \nu_r\,, \qquad \nu_L = 2\left(\nu_\phi - \nu_r\right)\,.
\label{eq:wd}
\end{equation}

\begin{figure*}[t]
\centering
\includegraphics[width=\linewidth]{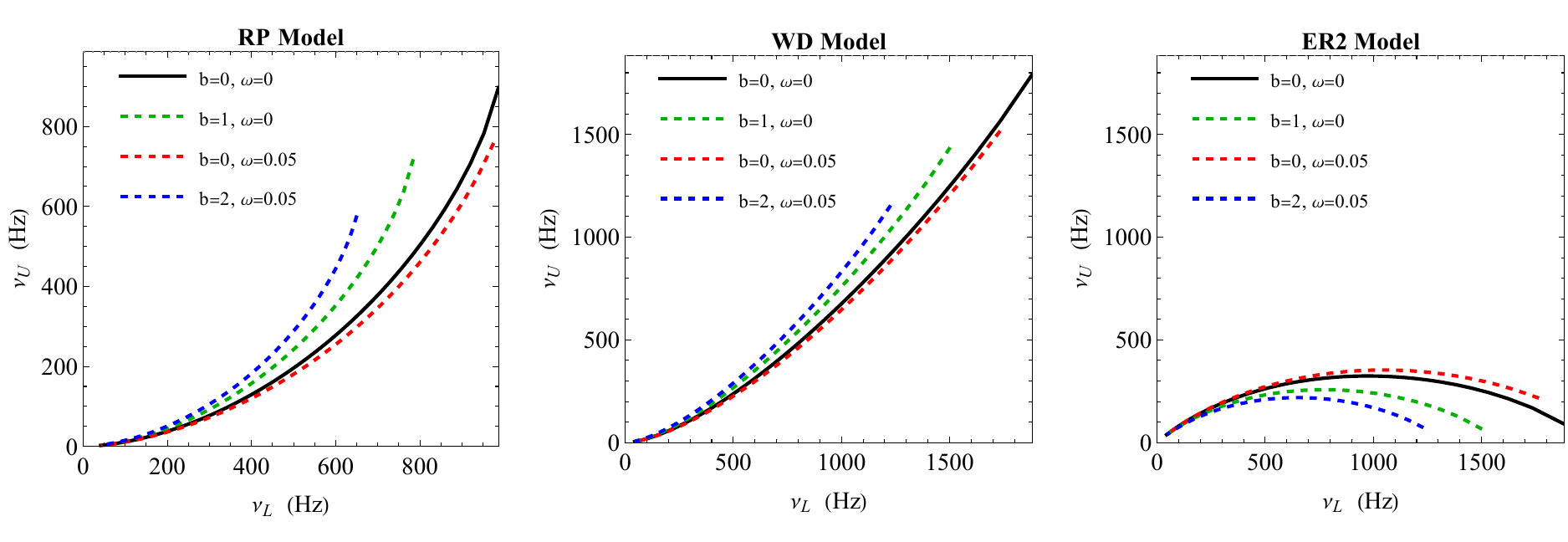}
\caption{Upper ($\nu_U$) versus lower ($\nu_L$) twin-peak QPO frequencies for the RP, WD, and ER2 models in the tidal charged braneworld background with the current-loop field, for selected values of $b$ and $\omega$.}
\label{fig:qpo_rp_wd_er2}
\end{figure*}

\begin{figure*}[t]
\centering
\includegraphics[width=\linewidth]{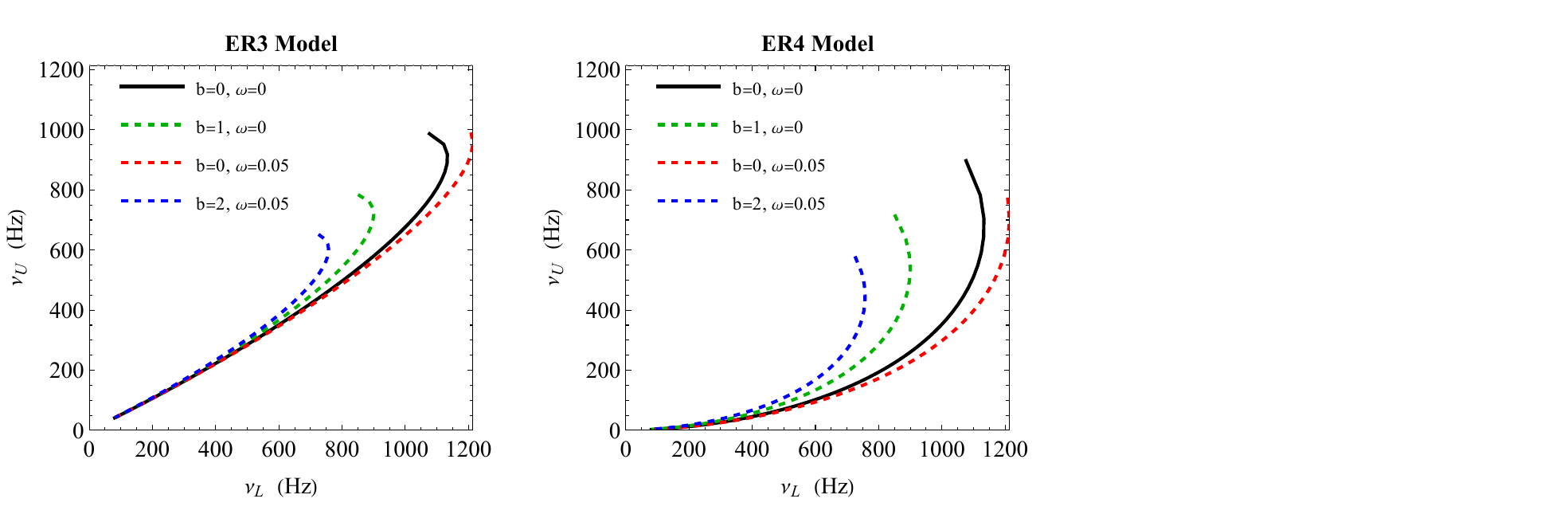}
\caption{Upper ($\nu_U$) versus lower ($\nu_L$) twin-peak QPO frequencies for the ER3 and ER4 models, for selected values of $b$ and $\omega$.}
\label{fig:qpo_er3_er4}
\end{figure*}

The $\nu_U$--$\nu_L$ correlations predicted by these five models are shown in
Figs.~\ref{fig:qpo_rp_wd_er2} and \ref{fig:qpo_er3_er4}, each curve generated by sweeping the emission radius. The locus $\nu_U=\nu_L$ marks the coalescence of the two peaks into a single feature. For the RP and WD models the tracks rise monotonically and steepen; increasing $b$ shifts them upward, raising $\nu_U$ at fixed $\nu_L$, following directly from the enhancement of the orbital frequency by the tidal charge (Sec.~\ref{sec:frequencies}), while $\omega$ lowers the curves only slightly. The ER2 model displays the characteristic rise-and-turnover ``arch,'' in which a larger $b$ lowers the arch while a positive $\omega$ raises it---an inversion relative to RP/WD that reflects the central role of the vertical epicyclic frequency. The ER3 and ER4 models develop the hooked profiles typical of these resonances. Overall, the tidal charge imprints a clear, systematic signature across all models, while the magnetic coupling acts as a finer, subdominant modulation most visible in the ER family, where it enters through $\nu_\theta$.

Evaluated under the observationally favored $3\!:\!2$ resonance, these twin-peak relations furnish the theoretical frequencies confronted with data in the next section.

\section{Monte Carlo Markov Chain (MCMC) analysis}
\label{sec:mcmc}

We now constrain the tidal charge $b$ and the magnetic coupling $\omega$ using the observed twin-peak QPO frequencies of six well-established black hole sources spanning three mass regimes: GRO~J1655$-$40, XTE~J1550$-$564, GRS~1915$+$105, H~1743$+$322, M82~X-1, and Sgr~A$^{*}$~\cite{Strohmayer2001,Remillard2002,Belloni2012,Motta2014}. The first four are stellar-mass black holes, M82~X-1 represents the intermediate-mass class, and Sgr~A$^{*}$ is supermassive. Their observational properties are summarized in Table~\ref{tab:data}. We adopt the epicyclic-resonance ER4 model as the representative framework for the corner plots, since it returns the most cleanly constrained posteriors, with magnetic-coupling distributions that are symmetric about zero and confined within the prior bounds; motivated by the observed $\nu_U/\nu_L\approx3\!:\!2$ locking, we adopt this ratio throughout~\cite{Bambi2014}.

The Bayesian posterior is
\begin{equation}
\mathcal{P}(\theta\,|\,D,\mathcal{M}) =
\frac{\mathcal{P}(D\,|\,\theta,\mathcal{M})\,\pi(\theta\,|\,\mathcal{M})}
{\mathcal{P}(D\,|\,\mathcal{M})}\,,
\label{eq:bayes}
\end{equation}
where $\pi(\theta)$ is the prior over $\theta=\{M,b,\omega,r/M\}$ and $\mathcal{P}(D\,|\,\theta,\mathcal{M})$ the likelihood. We assume Gaussian priors,
\begin{equation}
\pi(\theta_i)\propto
\exp\!\left[-\frac{1}{2}\left(\frac{\theta_i-\theta_{0,i}}{\sigma_i}\right)^{2}\right],
\quad \theta_{{\rm low},i}<\theta_i<\theta_{{\rm high},i}\,,
\label{eq:prior}
\end{equation}
with $\theta_{0,i},\sigma_i$ the literature mean and standard deviation and the bounds enforcing $b\ge0$ and a finite interval in $\omega$. The likelihood combines both peaks,
\begin{equation}
\log\mathcal{L}=\log\mathcal{L}_U+\log\mathcal{L}_L\,,
\label{eq:loglike}
\end{equation}
with
\begin{equation}
\log\mathcal{L}_{U,L} = -\frac{1}{2}\sum_{i}
\frac{\bigl(\nu_{U\!,L;i}^{\rm obs}-\nu_{U\!,L;i}^{\rm th}\bigr)^{2}}
{(\sigma_{U\!,L;i}^{\rm obs})^{2}}\,,
\label{eq:loglike_UL}
\end{equation}
where $\nu^{\rm obs}$ and $\nu^{\rm th}$ are the observed and theoretical upper/lower
frequencies for the $i$th source.

\begin{figure*}[t]
\centering
\includegraphics[width=0.48\textwidth]{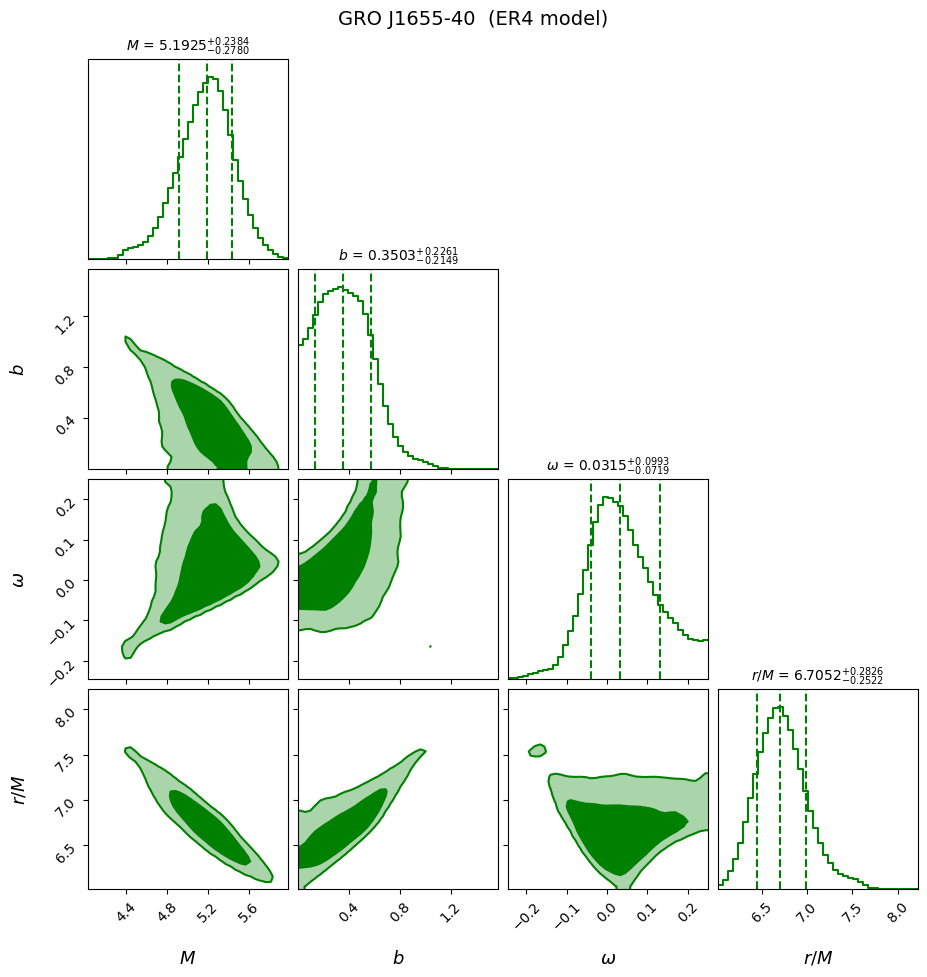}\hfill
\includegraphics[width=0.48\textwidth]{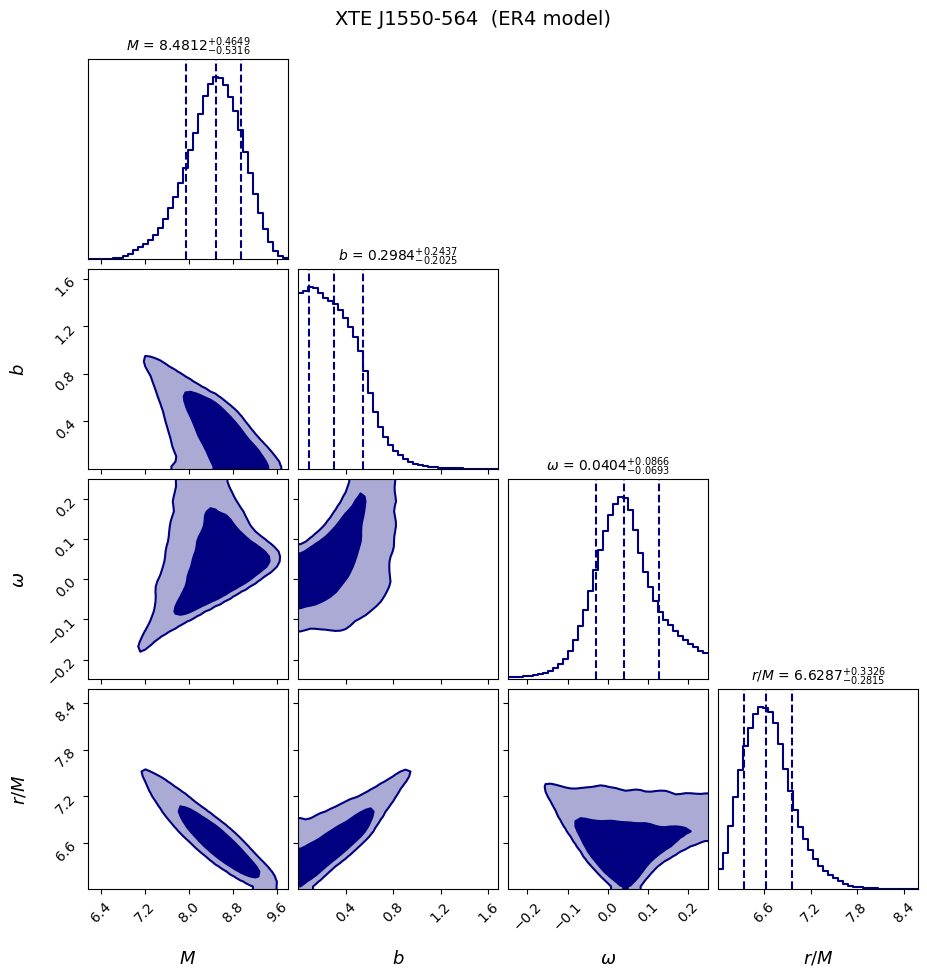}\\[6pt]
\includegraphics[width=0.48\textwidth]{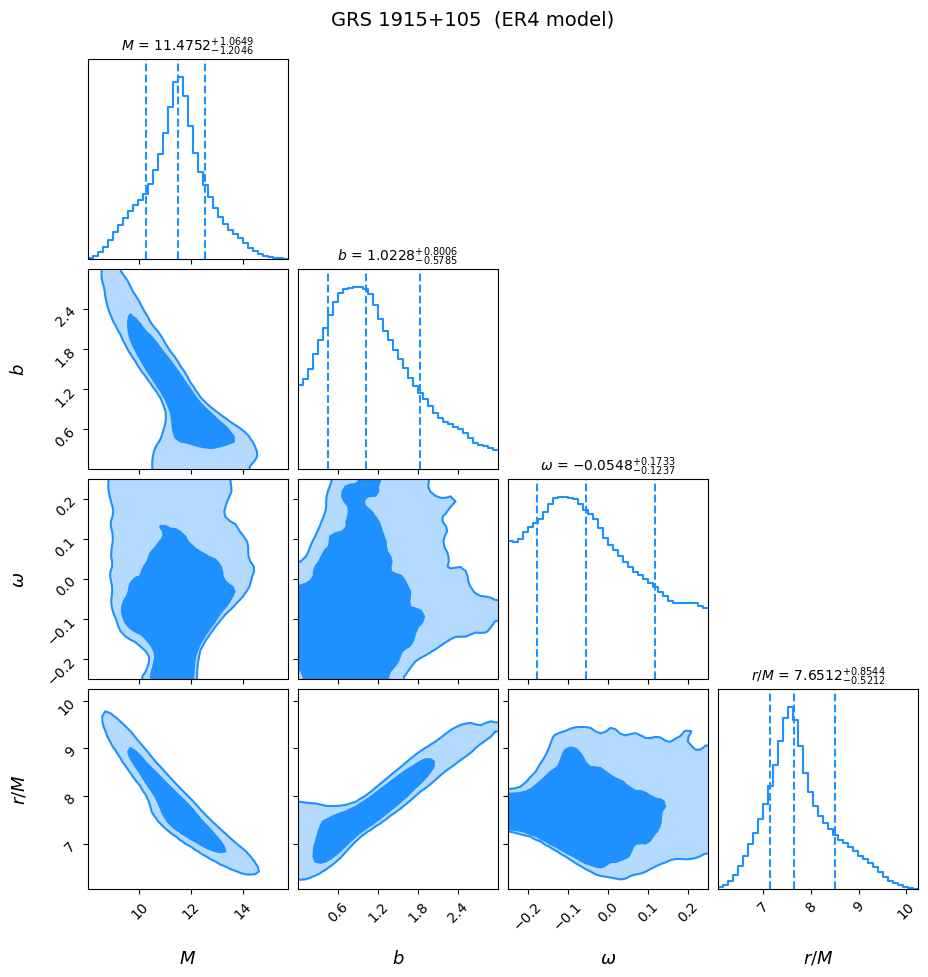}\hfill
\includegraphics[width=0.48\textwidth]{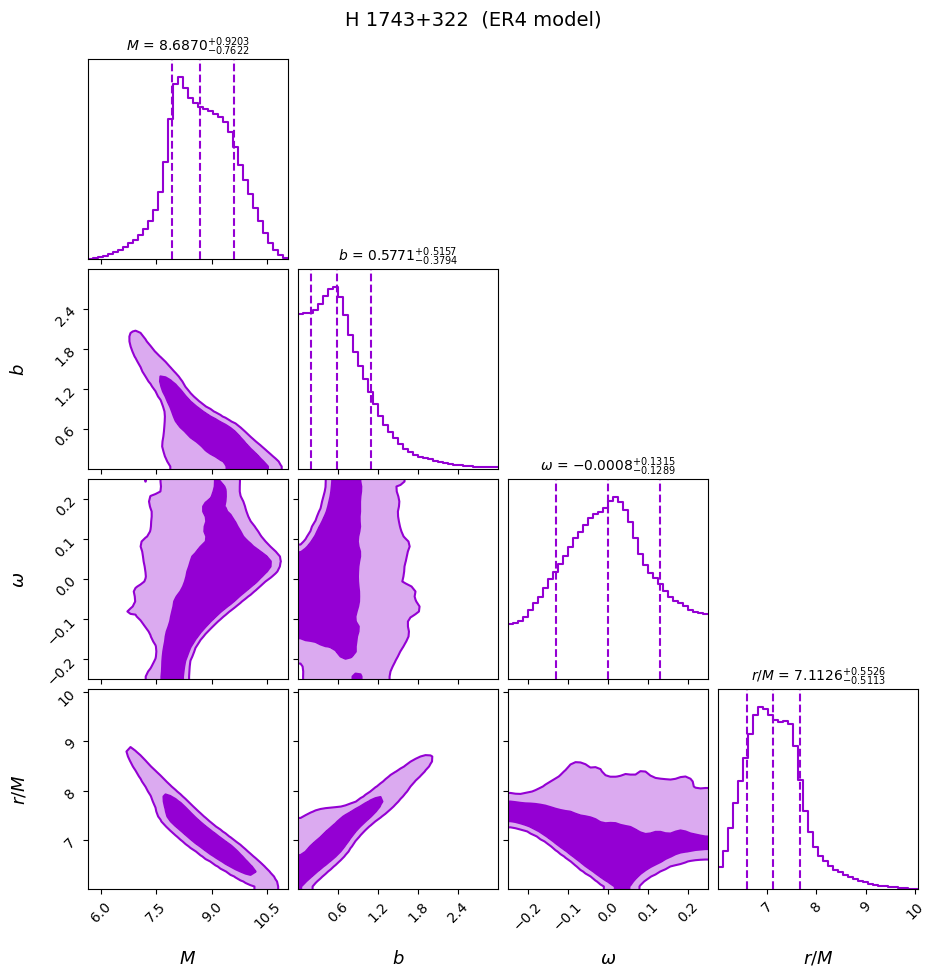}
\caption{Corner plots of the posterior distributions of $\{M,b,\omega,r/M\}$ from the MCMC analysis (ER4 model) for the stellar-mass sources GRO~J1655$-$40, XTE~J1550$-$564, GRS~1915$+$105, and H~1743$+$322. Shaded contours denote the $1\sigma$ and $2\sigma$ credible intervals.}
\label{fig:corner_stellar}
\end{figure*}

\begin{figure*}[t]
\centering
\includegraphics[width=0.48\textwidth]{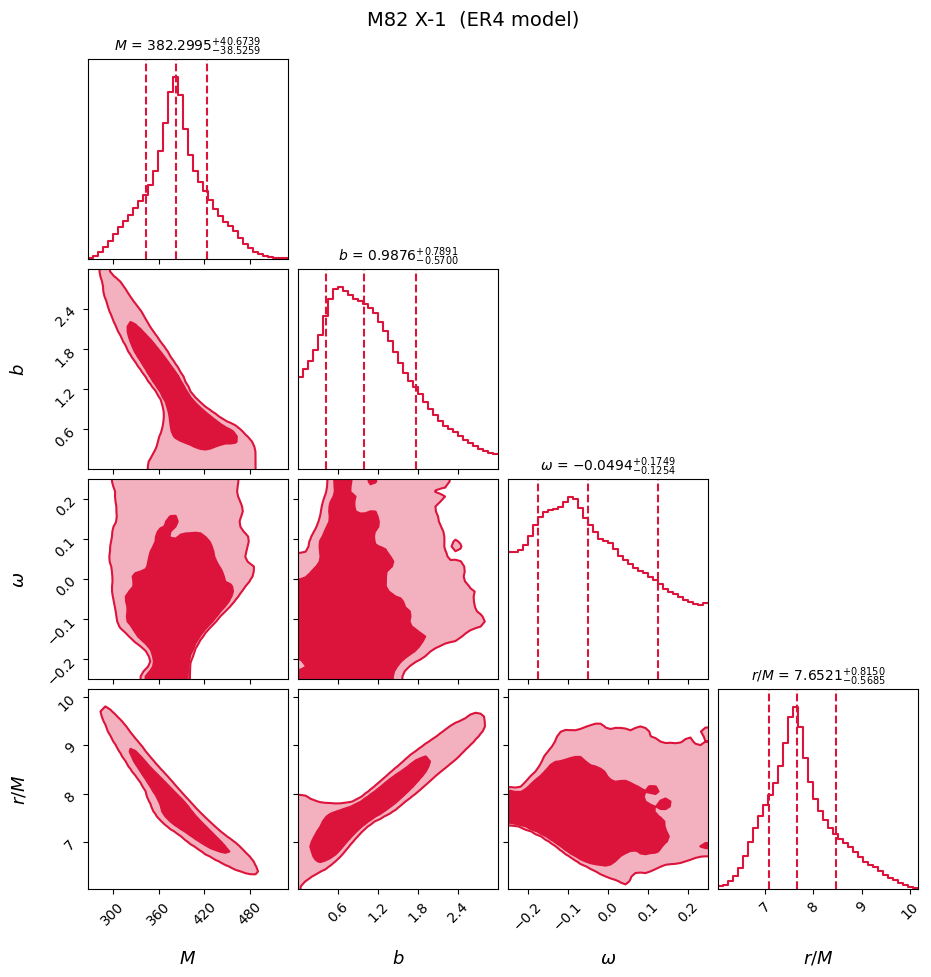}\hfill
\includegraphics[width=0.48\textwidth]{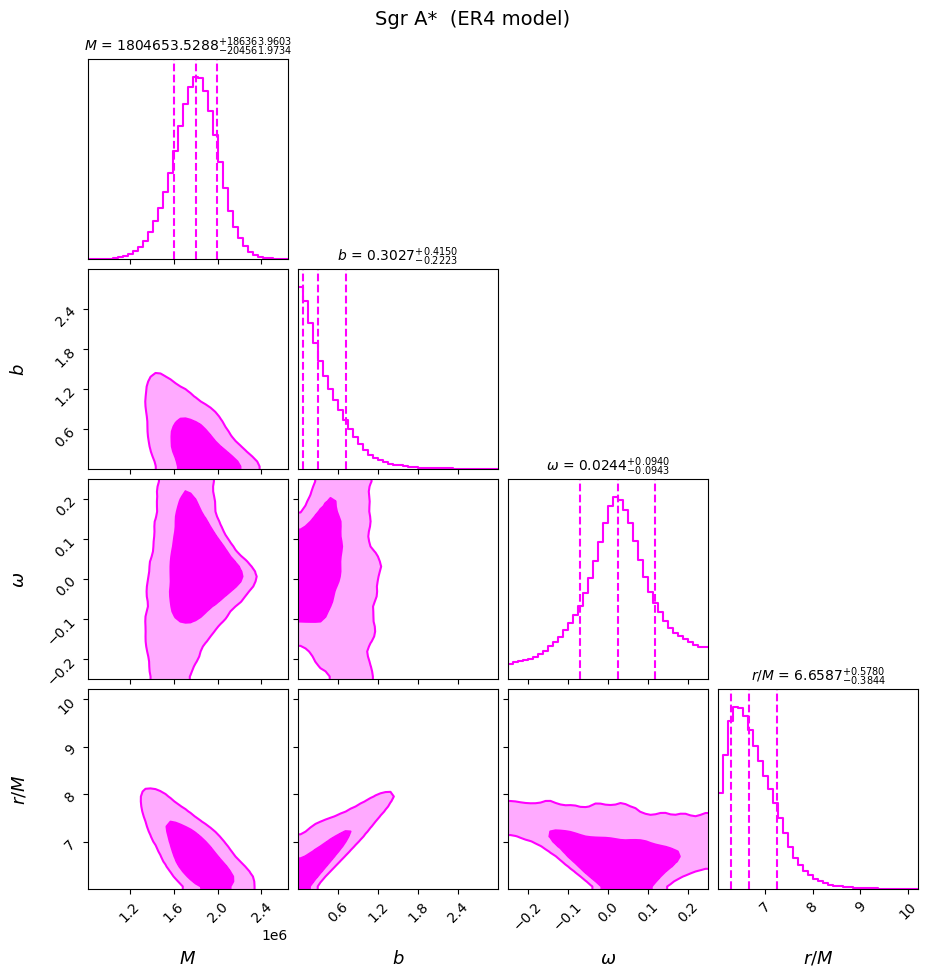}
\caption{Corner plots of the posterior distributions of $\{M,b,\omega,r/M\}$ from the MCMC analysis (ER4 model) for the intermediate-mass source M82~X-1 and the supermassive black hole Sgr~A$^{*}$.}
\label{fig:corner_massive}
\end{figure*}

We explore the posteriors with the affine-invariant ensemble sampler \texttt{emcee}~\cite{ForemanMackey2013}, running $6000$ steps per source; acceptance fractions range from $\sim0.32$ to $0.45$, indicating well-mixed chains. We determine the most probable $\{M,b,\omega,r/M\}$ across all five QPO models (RP, WD, ER2--ER4). Figures~\ref{fig:corner_stellar} and \ref{fig:corner_massive} show the ER4 corner plots, with $1\sigma$ and $2\sigma$ credible contours.

For the stellar-mass sources the inferred masses agree with the dynamical estimates: $M=5.19^{+0.24}_{-0.28}\,M_\odot$ (GRO~J1655$-$40), $8.48^{+0.46}_{-0.53}\,M_\odot$ (XTE~J1550$-$564), $11.48^{+1.06}_{-1.20}\,M_\odot$ (GRS~1915$+$105), and $8.69^{+0.92}_{-0.76}\,M_\odot$ (H~1743$+$322). The tidal charges lie in $b\simeq0.30$--$1.02$, largest for GRS~1915$+$105, while $\omega\simeq0.03,0.04,-0.055,-0.001$ is consistent with zero; the resonance radius is tightly clustered, $r/M\simeq6.63$--$7.65$.
The intermediate-mass source M82~X-1 gives $M=382.3^{+40.7}_{-38.5}\,M_\odot$, $b=0.99^{+0.79}_{-0.57}$, and $\omega=-0.049^{+0.175}_{-0.125}$ at $r/M=7.65^{+0.82}_{-0.57}$. For Sgr~A$^{*}$ we obtain $M=(1.805^{+0.19}_{-0.21})\times10^{6}\,M_\odot$, $b=0.30^{+0.42}_{-0.22}$, and
$\omega=0.024^{+0.094}_{-0.094}$ at $r/M=6.66^{+0.58}_{-0.38}$.

Across all mass regimes, the magnetic coupling $\omega$ remains small, with posteriors symmetric about the origin and consistent with zero, so that the data place an \emph{upper bound} on the electromagnetic interaction rather than a detection. The tidal charge $b$, by contrast, is driven to moderate, non-negligible values, $b\simeq0.30$--$1.02$ for ER4, and carries the dominant statistically meaningful signal in the QPO spectra. Both parameters remain compatible with the Schwarzschild limit $b\to0,\omega\to0$, so the tightly constrained posteriors support a small but measurable braneworld correction modulated by a weak current-loop field~\cite{Bambi2012,Johannsen2011,Bambi2014}.

Table~\ref{tab:fits} summarizes the best-fit parameters for all five models. The tidal charge shows moderate variation: ER4 returns $b\simeq0.30$--$1.02$, WD and ER3 span $b\simeq0.19$--$0.87$, RP favours smaller values $b\simeq0.07$--$0.52$, and ER2 yields the smallest, $b\lesssim0.15$. The magnetic coupling is smallest and most symmetric in ER4 ($|\omega|\lesssim0.05$), takes small positive values in RP, and is pushed toward the prior edge in WD/ER3 and, most markedly, in ER2. The recovered masses are closest to the dynamical estimates for WD and ER4, whereas ER2 systematically underestimates the masses and places the resonance at the largest radii, $r/M\simeq9.6$--$14.4$. Overall, ER4 combines faithful mass recovery with the tightest posteriors, motivating its choice as the reference framework; the consistency of small $\omega$ and moderate $b$ across all five models points to a universal, subtle imprint of the braneworld geometry and the current-loop field on the QPO structure of these black holes.

\begin{table*}[t]
\centering
\caption{Observational QPO data for the black hole sources used in the MCMC analysis, with independently estimated masses~\cite{Strohmayer2001,Remillard2002,Belloni2012,Motta2014}.}
\label{tab:data}
\begin{ruledtabular}
\begin{tabular}{lccc}
Source & Mass ($M_\odot$) & $\nu_U$ (Hz) & $\nu_L$ (Hz) \\
\colrule
GRO J1655$-$40 & $5.4\pm0.3$        & $441\pm2$   & $298\pm4$   \\
XTE J1550$-$564& $9.1\pm0.6$        & $276\pm3$   & $184\pm5$   \\
GRS 1915$+$105 & $12.4^{+2.0}_{-1.8}$& $168\pm3$  & $113\pm5$   \\
H 1743$+$322   & $8.0$--$14.0$      & $242\pm3$   & $166\pm5$   \\
M82 X-1        & $415\pm63$         & $5.07\pm0.06$ & $3.32\pm0.06$ \\
Sgr A$^{*}$    & $(3.5$--$4.9)\!\times\!10^{6}$ & $(1.445\pm0.16)\!\times\!10^{-3}$
               & $(0.886\pm0.04)\!\times\!10^{-3}$ \\
\end{tabular}
\end{ruledtabular}
\end{table*}

\begin{table*}[t]
\centering
\caption{Best-fit parameters (median with $68\%$ credible intervals) for the five QPO models across the six black hole sources.}
\label{tab:fits}
\begin{ruledtabular}
\begin{tabular}{llcccc}
Model & Source & $M\,(M_\odot)$ & $b$ & $\omega$ & $r/M$ \\
\colrule
RP & GRO J1655-40 & $4.6352^{+0.1158}_{-0.1541}$ & $0.0680^{+0.0983}_{-0.0508}$ & $0.1235^{+0.0365}_{-0.0320}$ & $6.2186^{+0.1747}_{-0.1221}$ \\
RP & XTE J1550-564 & $7.3607^{+0.2251}_{-0.2872}$ & $0.0816^{+0.1093}_{-0.0608}$ & $0.1275^{+0.0439}_{-0.0407}$ & $6.2616^{+0.2074}_{-0.1454}$ \\
RP & GRS 1915+105 & $9.6116^{+1.5480}_{-1.2861}$ & $0.5207^{+0.7735}_{-0.3589}$ & $0.0239^{+0.1357}_{-0.1633}$ & $7.5674^{+0.9479}_{-0.8686}$ \\
RP & H 1743+322 & $7.5387^{+0.7133}_{-1.0335}$ & $0.2752^{+0.4094}_{-0.1971}$ & $0.0791^{+0.0904}_{-0.1243}$ & $6.8447^{+0.8794}_{-0.4829}$ \\
RP & M82 X-1 & $320.93^{+50.10}_{-46.86}$ & $0.4418^{+0.6746}_{-0.3086}$ & $0.0439^{+0.1269}_{-0.1570}$ & $7.4780^{+1.0278}_{-0.8328}$ \\
RP & Sgr A* & $1.737^{+0.246}_{-0.260}\!\times\!10^{6}$ & $0.2985^{+0.4662}_{-0.2263}$ & $-0.0030^{+0.1013}_{-0.1137}$ & $6.7556^{+0.6087}_{-0.4524}$ \\
\colrule
WD & GRO J1655-40 & $5.2912^{+0.2970}_{-0.2997}$ & $0.2714^{+0.1843}_{-0.1685}$ & $0.1320^{+0.0783}_{-0.0737}$ & $7.5979^{+0.3637}_{-0.3226}$ \\
WD & XTE J1550-564 & $8.7675^{+0.5847}_{-0.5856}$ & $0.2039^{+0.1812}_{-0.1355}$ & $0.1512^{+0.0660}_{-0.0761}$ & $7.3765^{+0.4340}_{-0.3776}$ \\
WD & GRS 1915+105 & $10.8149^{+1.6810}_{-1.4832}$ & $0.8691^{+0.9295}_{-0.5736}$ & $-0.0066^{+0.1593}_{-0.1529}$ & $9.3770^{+1.0866}_{-1.0283}$ \\
WD & H 1743+322 & $9.3022^{+1.4112}_{-1.4547}$ & $0.3768^{+0.5233}_{-0.2648}$ & $0.0988^{+0.1033}_{-0.1408}$ & $7.9505^{+1.1523}_{-0.8477}$ \\
WD & M82 X-1 & $350.19^{+62.00}_{-51.00}$ & $0.6728^{+0.7574}_{-0.4493}$ & $0.0440^{+0.1351}_{-0.1542}$ & $9.2568^{+1.1930}_{-1.0937}$ \\
WD & Sgr A* & $3.228^{+0.417}_{-0.421}\!\times\!10^{6}$ & $0.4122^{+0.4927}_{-0.2965}$ & $0.0012^{+0.1288}_{-0.1420}$ & $6.9852^{+0.5428}_{-0.5310}$ \\
\colrule
ER2 & GRO J1655-40 & $2.3176^{+0.1328}_{-0.1279}$ & $0.0088^{+0.0145}_{-0.0066}$ & $0.2463^{+0.0028}_{-0.0062}$ & $11.4237^{+0.4064}_{-0.3839}$ \\
ER2 & XTE J1550-564 & $4.9910^{+0.3891}_{-0.3796}$ & $0.0122^{+0.0200}_{-0.0091}$ & $0.2445^{+0.0042}_{-0.0091}$ & $9.5683^{+0.4685}_{-0.4133}$ \\
ER2 & GRS 1915+105 & $5.2196^{+1.3213}_{-1.3293}$ & $0.1271^{+0.3049}_{-0.0992}$ & $0.2055^{+0.0333}_{-0.0949}$ & $12.7351^{+2.7173}_{-1.7093}$ \\
ER2 & H 1743+322 & $2.9433^{+0.6778}_{-0.6583}$ & $0.1226^{+0.3042}_{-0.0950}$ & $0.2042^{+0.0355}_{-0.1015}$ & $14.4248^{+2.6631}_{-1.8153}$ \\
ER2 & M82 X-1 & $149.33^{+27.95}_{-32.43}$ & $0.1492^{+0.3541}_{-0.1161}$ & $0.2007^{+0.0379}_{-0.1104}$ & $13.9435^{+2.5496}_{-1.5190}$ \\
ER2 & Sgr A* & $8.958^{+1.85}_{-2.41}\!\times\!10^{5}$ & $0.1015^{+0.2806}_{-0.0799}$ & $0.2110^{+0.0295}_{-0.0816}$ & $9.6505^{+2.4207}_{-1.2611}$ \\
\colrule
ER3 & GRO J1655-40 & $5.2879^{+0.2969}_{-0.2949}$ & $0.2686^{+0.1766}_{-0.1711}$ & $0.1328^{+0.0780}_{-0.0733}$ & $7.6033^{+0.3257}_{-0.2878}$ \\
ER3 & XTE J1550-564 & $8.7847^{+0.5938}_{-0.6037}$ & $0.1927^{+0.1717}_{-0.1302}$ & $0.1583^{+0.0613}_{-0.0717}$ & $7.4588^{+0.4034}_{-0.3527}$ \\
ER3 & GRS 1915+105 & $10.6841^{+1.7519}_{-1.4555}$ & $0.7804^{+0.8284}_{-0.5196}$ & $0.0332^{+0.1432}_{-0.1678}$ & $9.1742^{+1.1568}_{-0.9735}$ \\
ER3 & H 1743+322 & $9.2992^{+1.4453}_{-1.5173}$ & $0.3266^{+0.4660}_{-0.2324}$ & $0.1187^{+0.0890}_{-0.1335}$ & $7.7280^{+1.0626}_{-0.7525}$ \\
ER3 & M82 X-1 & $349.64^{+61.97}_{-53.21}$ & $0.6876^{+0.7692}_{-0.4613}$ & $0.0548^{+0.1266}_{-0.1579}$ & $9.4001^{+1.2432}_{-1.0377}$ \\
ER3 & Sgr A* & $2.262^{+0.196}_{-0.267}\!\times\!10^{6}$ & $0.1126^{+0.2090}_{-0.0850}$ & $0.1750^{+0.0550}_{-0.0996}$ & $6.5291^{+0.5794}_{-0.3108}$ \\
\colrule
ER4 & GRO J1655-40 & $5.1925^{+0.2384}_{-0.2780}$ & $0.3503^{+0.2261}_{-0.2149}$ & $0.0315^{+0.0993}_{-0.0719}$ & $6.7052^{+0.2826}_{-0.2522}$ \\
ER4 & XTE J1550-564 & $8.4812^{+0.4649}_{-0.5316}$ & $0.2984^{+0.2437}_{-0.2025}$ & $0.0404^{+0.0866}_{-0.0693}$ & $6.6287^{+0.3326}_{-0.2815}$ \\
ER4 & GRS 1915+105 & $11.4752^{+1.0649}_{-1.2046}$ & $1.0228^{+0.8006}_{-0.5785}$ & $-0.0548^{+0.1733}_{-0.1237}$ & $7.6512^{+0.8544}_{-0.5212}$ \\
ER4 & H 1743+322 & $8.6870^{+0.9203}_{-0.7622}$ & $0.5771^{+0.5157}_{-0.3794}$ & $-0.0008^{+0.1315}_{-0.1289}$ & $7.1126^{+0.5526}_{-0.5113}$ \\
ER4 & M82 X-1 & $382.30^{+40.67}_{-38.53}$ & $0.9876^{+0.7891}_{-0.5700}$ & $-0.0494^{+0.1749}_{-0.1254}$ & $7.6521^{+0.8150}_{-0.5685}$ \\
ER4 & Sgr A* & $1.805^{+0.186}_{-0.205}\!\times\!10^{6}$ & $0.3027^{+0.4150}_{-0.2223}$ & $0.0244^{+0.0940}_{-0.0943}$ & $6.6587^{+0.5780}_{-0.3844}$ \\
\end{tabular}
\end{ruledtabular}
\end{table*}

\clearpage
\section{Summary and conclusions}
\label{sec:conclusion}

We have studied the dynamics of charged particles in the dipolar electromagnetic field of a stationary current loop around a static tidal charged black hole in the Randall--Sundrum braneworld model~\cite{Randall1999,Dadhich2000}. Taking the exact azimuthal four-potential of Ref.~\cite{Turimov2018} as input and working entirely from the Hamilton--Jacobi equation, we derived the charged-particle effective potential in the interior and exterior regions of the loop and analyzed the circular orbits, ISCO, and fundamental frequencies as functions of the tidal charge $b$ and the magnetic coupling $\omega$. Our main results are as follows.

(i) The event horizon and the neutral ISCO both migrate outward as $b$ increases (Figs.~\ref{fig:horizon_b} and \ref{fig:isco_b}), reflecting the gravitationally attractive character of the negative tidal charge; the loop is anchored at $r_0=r_{\rm ISCO}(b)$, giving $r_0=6M,7.31M,8.38M$ for $b=0,1,2$.

(ii) The interior and exterior effective potentials match continuously at $r_0$ and reduce to the neutral case at $\omega=0$. Consequently, every ISCO quantity---$r_{\rm ISCO}$, $\mathcal{L}_{\rm ISCO}$, $\mathcal{E}_{\rm ISCO}$, and $\Omega_{\rm ISCO}$---returns to its neutral value at $\omega=0$, providing an exact cross-check on the interior and exterior branches.

(iii) A positive magnetic coupling raises the ISCO energy and can drive $\mathcal{E}_{\rm ISCO}>1$, unbinding the orbit and rendering the accretion efficiency $\eta=1-\mathcal{E}_{\rm ISCO}$ negative; we mapped the bound--unbound boundary in the $(b,\omega)$ plane (Figs.~\ref{fig:bound_in} and \ref{fig:bound_eisco}). Since the sign of $\omega$ tracks the sign of the charge, bound ISCO orbits favor positively charged particles in the loop field, with implications for the charge composition of accreting plasma~\cite{KhanChen2023,Rayimbaev2021}. The negative tidal charge lowers the ISCO binding energy relative to Schwarzschild, so it reduces the maximal radiative efficiency.

(iv) The $\sin^2\theta$ structure of the dipolar loop field splits the vertical and azimuthal frequencies [Eq.~\eqref{eq:omega_theta}], reviving a nonzero nodal precession frequency $\nu_\phi-\nu_\theta$ even though the background is static and non-rotating---a genuinely new feature relative to neutral geodesics that is controlled entirely by the product $\omega\,\mathcal{G}$~\cite{Stella1999,Tursunov2018}.

(v) Using these frequencies, we built five twin-peak QPO models and confronted them with the observed frequencies of six sources through an MCMC analysis~\cite{ForemanMackey2013}. Adopting ER4 as the representative framework, the tidal charge is driven to moderate values $b\simeq0.30$--$1.02$, while the magnetic coupling stays consistent with zero ($|\omega|\lesssim0.05$), symmetric about the origin. The data therefore place upper bounds on both the tidal charge and the magnetic coupling strength, and remain compatible with the Schwarzschild limit $b\to0,\omega\to0$~\cite{Bambi2012,Johannsen2011,Bambi2014}.

Taken together, these results show that a localized current-loop field around a tidal charged braneworld black hole produces a rich, region-dependent phenomenology in the ISCO properties, accretion efficiency, and QPO structure. The braneworld geometry provides the dominant, statistically meaningful signal, while the electromagnetic coupling enters as a weak, subdominant modulation. Extending the analysis to rotating tidal charged backgrounds and to a wider set of QPO sources would sharpen these constraints, and is left for future work.

\begin{acknowledgments}
\end{acknowledgments}


\begin{thebibliography}{99}

\bibitem{Petterson1974} J.~A.~Petterson, Phys. Rev. D \textbf{10}, 3166 (1974).
\bibitem{Petterson1975} J.~A.~Petterson, Phys. Rev. D \textbf{12}, 2218 (1975).
\bibitem{Wald1974} R.~M.~Wald, Phys. Rev. D \textbf{10}, 1680 (1974).
\bibitem{Piotrovich2010} M.~Y.~Piotrovich, N.~A.~Silant'ev, Y.~N.~Gnedin, and T.~M.~Natsvlishvili, arXiv:1002.4948.
\bibitem{Deutsch1955} A.~J.~Deutsch, Ann. Astrophys. \textbf{18}, 1 (1955).
\bibitem{Ginzburg1964} V.~L.~Ginzburg, Zh. Eksp. Teor. Fiz. \textbf{47}, 1030 (1964).
\bibitem{Rezzolla2001a} L.~Rezzolla, B.~J.~Ahmedov, and J.~C.~Miller, Mon. Not. R. Astron. Soc. \textbf{322}, 723 (2001).
\bibitem{Rezzolla2001b} L.~Rezzolla, B.~J.~Ahmedov, and J.~C.~Miller, Found. Phys. \textbf{31}, 1051 (2001).
\bibitem{Randall1999} L.~Randall and R.~Sundrum, Phys. Rev. Lett. \textbf{83}, 3370 (1999).
\bibitem{Dadhich2000} N.~Dadhich, R.~Maartens, P.~Papadopoulos, and V.~Rezania, Phys. Lett. B \textbf{487}, 1 (2000).
\bibitem{Turimov2018} B.~Turimov, Int. J. Mod. Phys. D \textbf{27}, 1850092 (2018).
\bibitem{Aliev2005} A.~N.~Aliev and A.~E.~G\"{u}mr\"{u}k\c{c}\"{u}o\u{g}lu, Phys. Rev. D \textbf{71}, 104027 (2005).
\bibitem{Abdujabbarov2010} A.~Abdujabbarov and B.~Ahmedov, Phys. Rev. D \textbf{81}, 044022 (2010).
\bibitem{Stuchlik2017} Z.~Stuchl\'{i}k, M.~Blaschke, and J.~Schee, Phys. Rev. D \textbf{96}, 104050 (2017).
\bibitem{Schee2009a} J.~Schee and Z.~Stuchl\'{i}k, Gen. Relativ. Gravit. \textbf{41}, 1795 (2009).
\bibitem{Schee2009b} J.~Schee and Z.~Stuchl\'{i}k, Int. J. Mod. Phys. D \textbf{18}, 983 (2009).
\bibitem{Turimov2017} B.~V.~Turimov, B.~J.~Ahmedov, and A.~A.~Hakimov, Phys. Rev. D \textbf{96}, 104001 (2017).
\bibitem{Ahmedov2008} B.~J.~Ahmedov and F.~J.~Fattoyev, Phys. Rev. D \textbf{78}, 047501 (2008).
\bibitem{Morozova2011} V.~S.~Morozova and B.~J.~Ahmedov, Astrophys. Space Sci. \textbf{333}, 133 (2011).
\bibitem{Stella1998} L.~Stella and M.~Vietri, Astrophys. J. Lett. \textbf{492}, L59 (1998).
\bibitem{Stella1999} L.~Stella and M.~Vietri, Phys. Rev. Lett. \textbf{82}, 17 (1999).
\bibitem{Bambi2012} C.~Bambi, Phys. Rev. D \textbf{85}, 043002 (2012).
\bibitem{Johannsen2011} T.~Johannsen and D.~Psaltis, Astrophys. J. \textbf{726}, 11 (2011).
\bibitem{Bambi2014} C.~Bambi, Phys. Rev. D \textbf{90}, 047503 (2014).
\bibitem{Strohmayer2001} T.~E.~Strohmayer, Astrophys. J. Lett. \textbf{552}, L49 (2001).
\bibitem{Remillard2002} R.~A.~Remillard, E.~H.~Morgan, J.~E.~McClintock, C.~D.~Bailyn, and J.~A.~Orosz, Astrophys. J. \textbf{580}, 1030 (2002).
\bibitem{Belloni2012} T.~M.~Belloni, A.~Sanna, and M.~M\'{e}ndez, Mon. Not. R. Astron. Soc. \textbf{426}, 1701 (2012).
\bibitem{ForemanMackey2013} D.~Foreman-Mackey, D.~W.~Hogg, D.~Lang, and J.~Goodman, Publ. Astron. Soc. Pac. \textbf{125}, 306 (2013).

\bibitem{AlZahrani2013} A.~M.~Al~Zahrani, V.~P.~Frolov, and A.~A.~Shoom, Phys. Rev. D \textbf{87}, 084043 (2013).
\bibitem{Igata2012} T.~Igata, T.~Harada, and M.~Kimura, Phys. Rev. D \textbf{85}, 104028 (2012).
\bibitem{Kopacek2014} O.~Kop\'{a}\v{c}ek and V.~Karas, Astrophys. J. \textbf{787}, 117 (2014).
\bibitem{Panis2019} R.~P\'{a}nis, M.~Kolo\v{s}, and Z.~Stuchl\'{i}k, Eur. Phys. J. C \textbf{79}, 479 (2019).
\bibitem{Stuchlik2020} Z.~Stuchl\'{i}k, M.~Kolo\v{s}, J.~Kov\'{a}\v{r}, P.~Slan\'{y}, and A.~Tursunov, Universe \textbf{6}, 26 (2020).
\bibitem{Tursunov2018} A.~Tursunov, M.~Kolo\v{s}, Z.~Stuchl\'{i}k, and D.~V.~Gal'tsov, Astrophys. J. \textbf{861}, 2 (2018).
\bibitem{Babar2016} G.~Z.~Babar, M.~Jamil, and Y.-K.~Lim, Int. J. Mod. Phys. D \textbf{25}, 1650024 (2016).
\bibitem{Turimov2020} B.~Turimov, J.~Rayimbaev, A.~Abdujabbarov, B.~Ahmedov, and Z.~Stuchl\'{i}k, Phys. Rev. D \textbf{102}, 064052 (2020).
\bibitem{Rayimbaev2021} J.~Rayimbaev, P.~Tadjimuratov, and A.~Abdujabbarov, Galaxies \textbf{9}, 75 (2021).
\bibitem{Murodov2023} S.~Murodov, J.~Rayimbaev, and B.~Ahmedov, Symmetry \textbf{15}, 2084 (2023).
\bibitem{Haydarov2020} K.~Haydarov, J.~Rayimbaev, A.~Abdujabbarov, S.~Palvanov, and D.~Begmatova, Eur. Phys. J. C \textbf{80}, 399 (2020).
\bibitem{KhanChen2023} S.~U.~Khan and Z.-M.~Chen, Eur. Phys. J. C \textbf{83}, 704 (2023).
\bibitem{Kurbonov2025} N.~Kurbonov, A.~H.~Bokhari, J.~Rayimbaev, and B.~Ahmedov, Eur. Phys. J. C \textbf{85}, 494 (2025).
\bibitem{AbramowiczKluzniak2001} M.~A.~Abramowicz and W.~Kluźniak, Astron. Astrophys. \textbf{374}, L19 (2001).
\bibitem{Torok2005} G.~T\"{o}r\"{o}k, M.~A.~Abramowicz, W.~Kluźniak, and Z.~Stuchl\'{i}k, Astron. Astrophys. \textbf{436}, 1 (2005).
\bibitem{Motta2014} S.~E.~Motta, T.~M.~Belloni, L.~Stella, T.~Mu\~{n}oz-Darias, and R.~Fender, Mon. Not. R. Astron. Soc. \textbf{437}, 2554 (2014).
\bibitem{IngramMotta2019} A.~R.~Ingram and S.~E.~Motta, New Astron. Rev. \textbf{85}, 101524 (2019).

\end{thebibliography}
\end{document}